\documentclass[letterpaper]{JHEP3}

\usepackage{amsfonts}
\usepackage{amssymb}
\usepackage{epsfig}
\usepackage{graphicx}
\usepackage{amsmath}
\usepackage{amsxtra}

\def\circa#1{\,\raise.3ex\hbox{$#1$\kern-.75em\lower1ex\hbox{$\sim$}}\,}
\makeatletter

\def\art{\@ifnextchar[{\eart}{\oart}}
\def\eart[#1]#2#3#4#5#6{{\rm #2}, {\em #3  #4} {\rm (#6) #5} ({\em #1})}
\def\hepart[#1]#2{{\rm #2, \em#1}}
\newcommand{\oart}[5]{{\rm #1}, {\em #2  #3} {\rm (#5) #4}}

\newcounter{alphaequation}[equation]
\def\thealphaequation{\theequation\hbox to
0.6em{\hfil\alph{alphaequation}\hfil}}
\def\eqnsystem#1{
\def\@eqnnum{{\rm (\thealphaequation)}}
\def\@@eqncr{\let\@tempa\relax \ifcase\@eqcnt \def\@tempa{& & &} \or
  \def\@tempa{& &}\or \def\@tempa{&}\fi\@tempa
  \if@eqnsw\@eqnnum\refstepcounter{alphaequation}\fi
\global\@eqnswtrue\global\@eqcnt=0\cr}
\refstepcounter{equation} \let\@currentlabel\theequation \def\@tempb{#1}
\ifx\@tempb\empty\else\label{#1}\fi
\refstepcounter{alphaequation}
\let\@currentlabel\thealphaequation
\global\@eqnswtrue\global\@eqcnt=0 \tabskip\@centering\let\\=\@eqncr
$$\halign to \displaywidth\bgroup \@eqnsel\hskip\@centering
$\displaystyle\tabskip\z@{##}$&\global\@eqcnt\@ne
\hskip2\arraycolsep\hfil${##}$\hfil& \global\@eqcnt\tw@\hskip2\arraycolsep
$\displaystyle\tabskip\z@{##}$\hfil
\tabskip\@centering&\llap{##}\tabskip\z@\cr}
\def\endeqnsystem{\@@eqncr\egroup$$\global\@ignoretrue} \makeatother

\def\be{\begin{equation}}
\def\ee{\end{equation}}
\def\bea{\begin{eqnarray}}
\def\eea{\end{eqnarray}}

\newcommand{\vo}{{\cal V}}

\newcommand{\roughly}[1]{\mathrel{\raise.3ex\hbox{$#1$\kern-0.85em
\lower1ex\hbox{$\sim$}}}}

\def\ba{\begin{eqnarray}}
\def\ea{\end{eqnarray}}
\def\be{\begin{equation}}
\def\ee{\end{equation}}

\def\NL{{\scriptscriptstyle NL}}

\def\O{\mathcal{O}}

\def\V{\mathcal{V}}

\def\nn{\nonumber}
\def\mc{\mathcal}

\def\({\left(}
\def\){\right)}

\def\pref#1{(\ref{#1})}

\title{Modulated Reheating and Large Non-Gaussianity\\ in String Cosmology}

\author{
M.~Cicoli,${}^{1,2}$
G.~Tasinato,${}^3$  I.~Zavala,${}^4$ C.P.~Burgess${}^{5,6}$ and F.~Quevedo${}^{1,7}$\\

$^1$ Abdus Salam ICTP, Strada Costiera 11, Trieste 34014, Italy \\
$^2$ INFN, Sezione di Trieste, Italy \\
$^3$ Institute of Cosmology \& Gravitation, University of Portsmouth,\\ Dennis Sciama Building, Portsmouth, PO1 3FX, UK\\
$^4$ Centre for Theoretical Physics, University of Groningen, \\Nijenborgh 4, 9747 AG Groningen, The Netherlands\\
$^5$ Department of Physics \& Astronomy, McMaster University,
 Hamilton ON, Canada\\
$^6$   Perimeter Institute for Theoretical Physics,
 Waterloo ON, Canada\\
$^7$ DAMTP/CMS, University of Cambridge, Wilberforce Road,
 Cambridge CB3 0WA, UK}

\abstract{A generic feature of the known string inflationary models is that the same physics that makes the inflaton lighter than the Hubble scale during inflation often also makes other scalars this light. These scalars can acquire isocurvature fluctuations during inflation, and given that their VEVs determine the mass spectrum and the coupling constants of the effective low-energy field theory, these fluctuations give rise to couplings and masses that are modulated from one Hubble patch to another. These seem just what is required to obtain primordial adiabatic fluctuations through conversion into density perturbations through the `modulation mechanism,' wherein reheating takes place with different efficiency in different regions of our Universe. Fluctuations generated in this way can generically produce non-gaussianity larger than obtained in single-field slow-roll inflation; potentially observable in the near future.
We provide here the first explicit example of the modulation mechanism at work in string cosmology, within the framework of LARGE Volume Type-IIB string flux compactifications. The inflationary dynamics involves two light K\"ahler moduli:
a fibre divisor plays the r\^{o}le of the inflaton whose decay rate to visible sector degrees of freedom
is modulated by the primordial fluctuations of a blow-up mode (which is made light by the use of poly-instanton corrections).
We find the challenges of embedding the mechanism into a concrete UV completion constrains the properties of the non-gaussianity that is found,
since for generic values of the underlying parameters, the model predicts a local bi-spectrum with $f_{\rm \NL}$ of order `a few'.
However, a moderate tuning of the parameters gives also rise
to explicit examples with $f_{\rm \NL}\sim \mc{O}(20)$ potentially observable by the Planck satellite.}

\preprint{DAMTP-2012-14}

\begin{document}

\section{Introduction}
\label{Introduction}

Although initially motivated as a solution to the initial-condition problems of standard Big-Bang cosmology, inflationary models \cite{Inflation} turn out also to provide a successful phenomenological description of the properties of primordial density fluctuations (which are predicted to arise as late-time consequences of quantum fluctuations of the inflaton).

Until relatively recently most of these predictions were made using single-field, slow-roll inflationary models, for which the inflaton is the only light scalar relevant during inflation \cite{SFRevs, SFeft}. The popularity of these models can be traced to several of their attractive features: ($i$)  {\em simplicity};  ($ii$) {\em effectiveness} (they describe primordial fluctuations very well); ($iii$) {\em predictivity} (near, but not exact, scale invariance, scale-independent spectral tilt, gaussian fluctuation statistics, and so on); and ($iv$) {\em robustness} (that is, predictions can be made without knowing the details of the cosmological history between the end of inflation and the beginning of the late-time Big-Bang epoch.)

\subsubsection*{Why defy Occam?}

Despite the simplicity and success of the single-field framework, two independent lines of reasoning suggest exploring multi-field models \cite{MFModels, etaproblem} in more detail. These two lines of reasoning have their roots both in observational developments as well as theoretical considerations, as we now describe.

The first reason to explore more complicated models arises from ongoing improvements in sensitivity of modern observations. In particular, besides testing the consistency of concordance cosmology through improved precision on cosmological parameters \cite{CosmoObs}, new instruments like the Planck satellite \cite{Planck} will soon probe the existence of non-gaussianity down to a level that can be produced by many types of multi-field models \cite{paul, wands} (though not by vanilla single-field slow-roll scenarios). It behooves us as a field to be alive to the kinds of physics to which these measurements might be sensitive, particularly if single-field slow-roll models were to be falsified by the discovery of non-gaussianity.

The second reason to explore more complicated models becomes evident once one tries to integrate inflationary cosmology into our broader understanding of physical law. This raises several independent issues. Most famously, inflationary models require the inflaton to be much lighter than the slowly changing Hubble scale during inflation, $m_\varphi \ll H$. This is a difficult thing to arrange once the inflaton is coupled to any other fields present at the energies of interest, even if these other fields do not play a direct r\^ole in cosmology \cite{etaproblem, InfNat}.

Furthermore, there are very likely to be a good number of other fields to which a putative inflaton might couple, although to make this precise requires specifying a concrete theory. The challenge when doing so is that we do not know the laws of physics at the high energies usually required \cite{MSSMInf}, however the proximity of $H$ to $M_P$ and the relevance of quantum fluctuations to primordial fluctuations strongly suggests that it should involve whatever makes sense of quantum gravity at high energies. Since string theory is presently our best-developed framework for understanding quantum gravity, it provides a natural framework for exploring these issues in a crisp way.

And it is in string cosmology that we meet the second reason for exploring multi-field inflationary models in more detail. This is because close to a decade of exploration reveals that multi-field dynamics is the norm for string inflation and not the exception \cite{StrInfRevs}. Although it remains difficult in string theory to achieve a sub-Hubble inflaton mass (and so also slow-roll inflation), once this has been done the same mechanism tends also to drag down the masses of several other fields as well, making them also potentially cosmologically active. This observation --- that multi-field models arise fairly generically in this way once embedded into microscopic models --- has recently motivated a more detailed study of the systematics of inflationary multi-field dynamics in general \cite{GenMF}.

\subsubsection*{Non-gaussianity and post-inflationary dynamics}

One of the main differences between single-field and multi-field inflationary models is the wide variety of fluctuation-generation mechanisms available in the multi-field case. Whereas single-field models must generate adiabatic curvature fluctuations, multi-field models can allow a variety of fluctuations in which the various fields need not alter the gravitational curvature. And although these isocurvature fluctuations would pose an observational problem for Cosmic Microwave Background (CMB) observations if they were to persist to the present epoch, they need not be bad during inflation if they can be converted to adiabatic fluctuations sometime between the epoch of horizon exit and more recent times. This could easily happen, such as if thermal equilibrium were established sufficiently early before the present epoch.

Indeed, generating adiabatic fluctuations through such a roundabout route might even be a virtue rather than a complication, since alternative mechanisms predict primordial fluctuations to depend differently on inflationary properties than the standard single-field mechanism. In particular, the interactions inherent amongst the fields in any multi-field mechanism can often generate observable amounts of non-gaussianity in the CMB, making any discovery of non-gaussianity into a new window into the physics responsible for generating fluctuations, and possibly into some aspects of the post-inflationary regime. Most interestingly, non-gaussianity produced in this way after inflation can often be distinguished observably from non-gaussianity generated directly during inflation, such as by non-standard single-field models \cite{SFeft,SFnongauss}.

The two best-developed mechanisms for converting, after inflation, inflationary isocurvature perturbations into late-epoch adiabatic fluctuations are the curvaton mechanism \cite{Curvaton} and modulation mechanism
\cite{DGZ1, Kofman, DGZ2, ZaldarriagaNG}. One might ask whether these ever arise in string-inflationary models, given the preponderance of multi-field scenarios these models have. Interestingly, they largely do not: for most string-inflation models explored to date only adiabatic fluctuations are appreciably generated at horizon exit, allowing their observational implications to be simply understood in terms of effective single-field models \cite{StrInfRevs, SIsinglefield}.

\subsubsection*{Isocurvature fluctuations from string vacua}

That is not to say that other means for generating fluctuations cannot arise in string theory, and ref.~\cite{BCGQTZ} provides a construction of an inflationary cosmology that uses the curvaton mechanism, within an explicit string vacuum in the LARGE Volume Scenario (LVS) \cite{LV, astro, GenAnalofLVS}. What is perhaps remarkable is that no instances of the modulation mechanism have so far arisen, given that this mechanism was designed with string theory in mind. (The modulation mechanism starts with the observation that couplings in string theory are often given as the values of fields, and so can vary from one Hubble patch to another if these fields so vary. If, in particular, the couplings that vary this way are involved in the transfer of energy from the inflaton into the heat of the later hot Big-Bang epoch, then this reheating would be modulated from one Hubble patch to another, in a way that might describe the observed primordial fluctuations. In particular, such a modulation has been argued generically to produce non-gaussianity at the soon-to-be observable level of $f_{\rm \NL} \sim 20$ \cite{ZaldarriagaNG}.)

The goal of this paper is to provide an explicit string realisation of this modulation mechanism, much as ref.~\cite{BCGQTZ} did for the curvaton mechanism. We are able to do so, but it proves to be unexpectedly difficult to do, for a variety of reasons. The main reason for this is the difficulty in assembling in one model all of the things that are assumed when building a modulated reheating model: the requirement that inflation takes place in the first place; the presence of a modulating field that is light during inflation, modulates the inflaton decay rate and is almost decoupled from the inflaton; and the conditions required for the modulation mechanism to beat the standard one. Finally, we ask that the modulating field decays before the inflaton, ensuring the absence of any curvaton-like contributions to the generation of the density perturbations.

Here we provide the first explicit string model for which all these conditions apply. By doing so within the LVS we also avoid the need for fine-tuning. What must be chosen carefully, however, is the underlying extra-dimensional manifold, including a particular brane set-up and a particular choice of world-volume fluxes. In particular we make use of novel ingredients in string model building, such as poly-instanton \cite{PolyInst} contributions to the superpotential, that allow the exploration of new kinds of phenomenology and cosmology within the LVS \cite{StringyADD,PolyInfl}.

LVS string vacua have several features that make them particularly attractive from the point of view of model building in cosmology. One of these is what makes the LVS attractive for other applications as well: its explicit incorporation of moduli stabilisation within a systematic expansion of the low-energy potential in inverse powers of the extra-dimensional volume. That is, if the extra-dimensional volume (in string units) is $\V = M_s^6 \Omega_6$, then the LVS scenario starts with the observation that for large $\V$ the low-energy scalar potential has the generic form of a sum of terms of order $1/\V^3$, $e^{-a \tau}/\V^2$ and $e^{-2a \tau}/\V$, where $\tau \simeq \O(10)$ is the dimensionless volume, $\tau = M_s^4 \Omega_4$, of a comparatively small four-cycle in the geometry. For some choices of parameters this has a minimum with $\V \simeq e^{a \tau}$, which makes very large volumes natural to obtain.

The low-energy potential, $V$, has another property that is of more specific interest for inflationary applications, however. This is that the leading, $\O(\V^{-3})$, contributions to the potential can be independent of many of the other extra-dimensional moduli, $\phi^i$. These moduli therefore can only enter $V$ at subdominant order in $1/\V$, in practice arising at one string loop \cite{GenAnalofLVS}:
\be
 V \simeq \frac{ f_0(\V e^{-a \tau})}{\V^3} +
 \frac{f_1(\phi^i)}{\V^{10/3}} + \cdots \,,
\ee
where $f_0$ and $f_1$ are calculable functions. For large $\V$ this allows these moduli to be parametrically light compared with others \cite{ubernat}, and in particular to be small compared with the Hubble scale, $H \simeq \sqrt{V}/M_P \simeq \O(\V^{-3/2})$. This both explains why the inflaton is light \cite{fiberinfl} and why many other fields are also equally light \cite{BCGQTZ}.

In the particular construction explored here, inflation is driven naturally by a fibre modulus --- similar to \cite{fiberinfl} --- which acquires a potential from loop corrections, as above.
A particular blow-up mode also remains light during inflation since it is stabilised as in \cite{StringyADD} due to subleading
poly-instanton corrections \cite{PolyInst}. The absence of single-instanton contributions to the superpotential
for this rigid four-cycle is guaranteed by the presence of extra fermionic zero modes which are Wilson line modulini.
Because this blow-up mode is lighter than the Hubble parameter during inflation, it acquires significant isocurvature fluctuations.

The key observation is that this field can play the r\^ole of a modulation field because its VEV controls the size of the inflaton coupling to visible sector degrees of freedom, resulting in a modulated decay rate due to the fluctuations of the blow-up mode. Reheating takes place as in \cite{KLS97, WReheat, Reheating} due to the perturbative decay of the inflaton, converting the initial isocurvature perturbations of the light blow-up mode into adiabatic perturbations that swamp the standard contributions coming from inflationary fluctuations of the inflaton field itself. Finally, couplings of the modulating field also ensure that it decays before the inflaton, ensuring the absence of curvaton-like contributions to primordial density perturbations.

\subsubsection*{Predictions for non-gaussianity}

There are several advantages to having such an explicit UV completion of the modulation mechanism. First, as is often the case, string theory provides a transparent geometrical interpretation of the modulation mechanism in terms of the properties of the internal manifold under consideration. Second, the consistency of the construction sets strong constraints on the final prediction for the non-gaussianities, whose properties we analyse in some detail. In particular, we find non-gaussianities of local form with a bi-spectrum parameter, $f_{\rm \NL}\sim \mc{O}(20)$. Although this is potentially observable by the Planck satellite, we find small tri-spectrum parameters, $g_{\rm \NL}\sim \mc{O}(500)$ and $\tau_{\rm \NL}\sim \mc{O}(1000)$.\footnote{Such values of tri-spectrum parameters  might however be observable by future CMB experiments as EPIC \cite{EPIC} or by large scale structure surveys \cite{NGreview}.}

The paper is organised as follows. \S2\ briefly reviews the framework of LVS compactifications on K3 or $T^4$ fibred Calabi-Yau three-folds. With later applications to the modulation mechanism in mind, we choose a manifold with five moduli: the fibre divisor, $\tau_1$ (which plays the r\^{o}le of the inflaton); the modulus of the base of the fibration, $\tau_2$ (which controls the overall volume $\vo$);
the first blow-up mode, $\tau_3$ (whose non-perturbative effects fix the volume $\V$ at exponentially large values);
a second blow-up mode, $\tau_4$ (which supports the visible sector);
and a third blow-up mode, $\tau_5$ (which intersects $\tau_4$ and plays the r\^ole of the modulating field).\footnote{Our notation follows
ref.~\cite{fiberinfl} as closely as possible.}
We seek a regime with moduli stabilised with the hierarchy $\tau_2 \sim \tau_1 \gg \tau_3 \sim \tau_4 \sim \tau_5$ which, because of the LV `magic,' can be done using hierarchies among the input fluxes that are at most $\O(10)$.

\S3\ then describes the inflationary scenario, including a cartoon of how reheating takes place, in order to set up a viable modulation mechanism. In \S4\ we then describe how primordial fluctuations are generated by the post-inflationary dynamics,
showing that the $\V$-dependence of the masses and couplings can produce acceptable adiabatic fluctuations from the modulation mechanism. Our conclusions are briefly summarised in \S5.

\section{The LARGE-volume string inflationary model}
\label{sec-consideration}

In this section, we identify  a set of moduli fields with potentially interesting cosmological applications. We derive the leading order contributions to the potential and kinetic terms for these fields, to be used  in the next sections for developing a model of inflation accompanied by modulated reheating. We will keep concise in this section, since we mainly apply techniques developed in other works to which we refer.

\subsection{Field content}
\label{sec-fieldcont}

The model we develop requires a compactification based on a
Calabi-Yau manifold with a K3 or $T^4$-fibred structure, characterised by at least five K\"ahler moduli, two of which are light during
inflation and so are relevant for cosmology. We list them here:
\begin{itemize}
\item[i)] A fibre modulus, $\tau_1$, playing the r\^ole of inflaton field as in \cite{fiberinfl} and wrapped by a stack of D7-branes.\footnote{These D7-branes support a hidden sector since this cycle is non-rigid and tends to be stabilised rather large in string units yielding an exponentially small gauge coupling.} The low-energy scalar potential acquires a dependence on $\tau_1$ through string loop contributions sourced by the D7s \cite{GenAnalofLVS,stringloops}. The potential is naturally flat enough to drive inflation.

\item[ii)] A base modulus, $\tau_2$, that mainly controls the overall extra-dimensional volume  stabilised at exponentially large values. It is wrapped by the same stack of D7-branes needed to generate the string loop potential for $\tau_1$.$^3$ This modulus is heavy during inflation, and lies on its minimum throughout this phase.

\item[iii)] A blow-up mode $\tau_3$,  an assisting field
    required to stabilise the volume ${\cal V}$ at its minimum. Its presence is usual in LARGE volume set-ups. The potential depends on $\tau_3$ through non-perturbative contributions generated by a stack of D7-branes wrapping $\tau_3$ and supporting a hidden sector that undergoes double gaugino condensation in a race-track way.\footnote{This blow-up mode cannot support a visible sector due to the tension between non-perturbative effects and chirality \cite{blumen}.} It is heavy during inflation.

\item[iv)] Two intersecting blow-up modes, $\tau_4$ and $\tau_5$.
$\tau_4$ supports a GUT or MSSM-like construction with D7-branes and it
is stabilised in the geometric regime by $D$-terms, which render this field heavy during inflation.
On the other hand $\tau_5$ is light during inflation,
and plays the r\^ole of modulating field in our cosmological application. In fact, $\tau_5$ is wrapped by an Euclidean D3-brane (E3) instanton
which generates tiny poly-instantons corrections to the superpotential that fix this modulus at subleading order.
Single instanton contributions associated with $\tau_5$ are absent due to the presence of Wilson line modulini which
give rise to extra fermionic zero modes.
\end{itemize}

\subsection{Compactification}
\label{compactification}

In order to realise our scenario, we consider an orientifold of a Calabi-Yau three-fold $X$
with a K3 or a $T^4$ fibration structure and $h^{1,1}=5$.\footnote{We focus on orientifold involutions such that $h^{1,1}_-=0$, so that $h^{1,1}_+=h^{1,1}$.}
Hence we have five smooth divisors $D_i$, $i=1,...,5$,
whose volumes are given by the real part of the K\"{a}hler moduli $\tau_i= {\rm Vol} (D_i)$ .
The K\"{a}hler form $J$ can be expanded in a basis of dual two-forms as\footnote{We expanded the K\"{a}hler form $J$ taking a minus sign for $\hat{D}_i$, $i=3,4,5$ in order to have positive two-cycle volumes also for blow-up modes.} $J= t_1 \hat{D}_1 +t_2 \hat{D}_2-t_3 \hat{D}_3-t_4 \hat{D}_4-t_5 \hat{D}_5$.

The volume is expressed in terms of the two-cycle moduli as \cite{K3fibCY,CMV}:
\be
 \vo= \frac{1}{6}\int_X J\wedge J\wedge J =
 {\mathbf a}_1 t_1 t_2^2 - \mathbf a_3 t_3^3 - (\mathbf a_4 t_4 + \mathbf a_5 t_5)^3 - \mathbf b_4 t_4^3,
\label{volts}
\ee
where ${\mathbf a}_i>0$ $\forall i=1,...,5$ and $\mathbf b_4>0$ given that every Calabi-Yau manifold is characterised by the fact that the signature of the matrix $\frac{\partial^2 V}{\partial t_i \partial t_j}$ is $(1, h^{1,1}-1)$ \cite{CDLO}, so with 1 positive and 4 negative signs in our case. The triple intersection numbers are given by \footnote{In the absence of singularities, the triple intersection numbers $k_{ijk} \in \mathbb{Z}$.}:
\be
 k_{ijk}=\int_X \hat{D}_i\wedge\hat{D}_j\wedge\hat{D}_k,
\ee
and are related to the parameters in (\ref{volts}) as follows:
$k_{122}= 2\, \mathbf a_1$, $k_{333}= 6 \,\mathbf a_3$, $k_{444}=6 \,\left(\mathbf a_4^3+ \mathbf b_4\right)$, $k_{445}=6 \, \mathbf a_4^2 \,\mathbf a_5$, $k_{455}= 6 \,\mathbf a_4\, \mathbf a_5^2$, $k_{555}=6 \,\mathbf a_5^3$.

This particular structure of the intersection numbers implies that the $D_1$ is
a K3 or $T^4$ divisor fibred over a $\mathbb{P}^1$ base whose volume is $t_1$,
$D_3$ is a diagonal del Pezzo divisor,
whereas $D_4$ and $D_5$ are two rigid four-cycles intersecting over a two-cycle given by:
\be
 {\rm Vol}\left(D_4\cap D_5\right)=\int_X J\wedge \hat{D}_4 \wedge \hat{D}_5 = - \left( k_{445} t_4 + k_{455} t_5\right)
 = - 6 \,\mathbf a_4 \mathbf a_5 \left( \mathbf a_4 t_4 + \mathbf a_5 t_5\right).
\label{intersection}
\ee
We shall also assume that $D_5$ is a rigid four-cycle with Wilson lines, {\em i.e.}~$h^{2,0}(D_5)=0$ while $h^{1,0}(D_5)\neq 0$
(for explicit examples of this kind of divisors in this context see \cite{K3fibCY}).

The four-cycle moduli are defined as:
\be
 \tau_i =\frac{1}{2}\int_X J\wedge J\wedge \hat{D}_i,
\ee
and so they take the form:
\begin{eqnarray}
 \tau_1 &=& \frac{1}{2}k_{122}t_2^2 = \mathbf a_1 t_2^2, \text{ \ \ \ \ }
 \tau_2 = k_{122}t_1 t_2 =  2 \mathbf  a_1 t_1 t_2, \text{ \ \ \ \ }
 \tau_3 = \frac{1}{2}k_{333}t_3^2 = 3 \mathbf a_3 t_3^2, \\
 \tau_4 &=& \frac{1}{2}  \left(k_{444} t_4^2+ 2 k_{445} t_4 t_5 +k_{455} t_5^2\right) =
 3 \mathbf a_4 \left(\mathbf a_4 t_4+ \mathbf a_5 t_5\right)^2+3 \mathbf b_4 t_4^2, \\
 \tau_5 &=& \frac{1}{2}  \left(k_{445} t_4^2+ 2 k_{554} t_4 t_5 + k_{555} t_5^2\right) =
 3 \mathbf a_5  \left(\mathbf a_4 t_4 + \mathbf a_5 t_5\right)^2.
\end{eqnarray}
It is convenient to rewrite the volume in terms of the four-cycle moduli as:
\be
 \vo\,=\,\alpha \left[ \sqrt{\tau_1} \tau_2-\gamma_3 \tau_3^{3/2}- \gamma_5 \tau_5^{3/2}
 -\gamma_4 \left(\tau_4- x\,\tau_5\right)^{3/2} \right], \label{initialVolume}
\ee
where we set $\alpha=1/\sqrt{4 \mathbf a_1}$, $\gamma_3= \sqrt{4 \mathbf a_1/(27 \mathbf a_3)}$, $\gamma_4 =  \sqrt{4
\mathbf a_1/(27 \mathbf b_4)}$,
$\gamma_5=  \sqrt{4 \mathbf a_1/(27 \mathbf a_5^3)}$ and $x=\mathbf a_4/\mathbf a_5$.

\subsection{Brane set-up and fluxes}
\label{set-up}

In this section we shall perform an explicit choice of brane set-up and fluxes that can
give rise to the desired phenomenological features of a modulated reheating scenario.
We consider the visible sector to be realised by two stacks of D7-branes
wrapped around the divisors $D_{\rm vis}$ and $D_{\rm int}$ \footnote{The suffix `vis' stays for `visible'
whereas `int' stays for `intersecting' since we have in mind a GUT-like theory living on $D_{\rm vis}$
with chiral matter arising from an intersection with the stack on $D_{\rm int}$. However this set-up contains also
the main features of MSSM-like constructions with the visible sector located on both divisors.}
together with an E3 instanton wrapped around $D_{\rm E3}$ \footnote{We neglect the hidden sector D7-branes wrapped around
$D_1$ and $D_2$ (in order to yield the loop generated inflationary potential) and $D_3$ (in order to
yield the racetrack superpotential) since they do not play any r\^ole in the modulation mechanism.}:
\be
D_{\rm vis} = v_4 D_4+ v_5 D_5\,, \quad D_{\rm E3} = D_5\,, \quad D_{\rm int} = i_4 D_4 + i_5 D_5\,.
\ee
We also turn on the following world-volume fluxes:
\be
F_{\rm vis} = f_4 \hat D_4+ f_5 \hat D_5 +\frac 12 \hat D_{\rm vis} \quad F_{\rm E3} = \frac 12 \hat D_{\rm E3},
\quad F_{\rm int} = g_4 \hat D_4 + g_5 \hat D_5 + \frac12 \hat D_{\rm int}\,,
\ee
where the half-integer contributions come from the cancellation of the Freed-Witten anomalies.
Let us fix the $B$-field in order to cancel the total flux $\mc{F}=F-B$ on $D_{\rm E3}$ so that
the instanton contributes to the superpotential.
Hence we choose $B=F_{\rm E3}$ which gives rise to the following total world-volume fluxes: \footnote{Even though we are focusing on orientifold involutions such that $h^{1,1}_-=0$, the $B$-field is not completely projected out, since it can have non-zero discrete coefficients which can compensate the half-integer contribution in $F_{E3}$.}
\be
\mc{F}_{\rm vis} =\left( f_4 +\frac{v_4}{2}\right) \hat D_4+ \left(f_5 +\frac{v_5-1}{2}\right) \hat D_5,
\quad \mc{F}_{\rm E3} = 0,
\quad \mc{F}_{\rm int} = \left( g_4 +\frac{i_4}{2}\right) \hat D_4+ \left(g_5 +\frac{i_5-1}{2}\right) \hat D_5. \notag
\ee
The presence of non-vanishing gauge fluxes has four implications:
\begin{enumerate}
\item They induce $U(1)$-charges for the K\"ahler moduli. The charges of the blow-up moduli
$T_4$ and $T_5$ under the diagonal $U(1)$ of the D7-stacks wrapping $D_{\rm vis}$ and $D_{\rm int}$ read:
\be
q_{{\rm vis}\,i}=\int_X \mc{F}_{\rm vis}\wedge \hat D_{\rm vis} \wedge \hat D_i, \qquad
q_{{\rm int}\,i}=\int_X \mc{F}_{\rm int}\wedge \hat D_{\rm int} \wedge \hat D_i, \qquad i=4,5\,.
\ee
These charges can be written one in terms of the other as:
\be
q_{{\rm vis}\,4}=x\, q_{{\rm vis}\,5} + 6 {\mathbf b}_4 v_4 \left(f_4 +\frac{v_4}{2}\right), \qquad
q_{{\rm int}\,4}=x\, q_{{\rm int}\,5} + 6 {\mathbf b}_4 i_4 \left(g_4 +\frac{i_4}{2}\right).
\label{chargerelations}
\ee

\item The gauge coupling $g_i$ of the field theory on the D7-stack wrapping the divisor $D_i$ acquires a flux-dependent shift. In fact,
 the gauge coupling is given by:
\be
4 \pi g_i^{-2} = \tau_i - h (\mc{F}_i)  \text{Re}(S)\,,
\label{gaugecoupling}
\ee
where $\text{Re}(S)=e^{-\phi}=g_s^{-1}$ is the real part of the axio-dilaton, while the flux-dependent factor is:
\bea
h (\mc{F}_{\rm vis})&=&\frac 12 \int_X \mc{F}_{\rm vis}\wedge \mc{F}_{\rm vis}\wedge \hat D_{\rm vis}
=\frac 12 \left[\left( f_4 +\frac{v_4}{2}\right)q_{{\rm vis}\,4}+\left(f_5 +\frac{v_5-1}{2}\right)q_{{\rm vis}\,5}\right], \nonumber \\
h (\mc{F}_{\rm int})&=&\frac 12 \int_X \mc{F}_{\rm int}\wedge \mc{F}_{\rm int}\wedge \hat D_{\rm int}
= \frac 12 \left[\left( g_4 +\frac{i_4}{2}\right)q_{{\rm int}\,4}+\left(g_5 +\frac{i_5-1}{2}\right)q_{{\rm int}\,5}\right]. \nonumber
\eea

\item The world-volume fluxes generate moduli-dependent Fayet-Iliopoulos (FI) terms which take the form:
\bea
 \xi_{\rm vis}&=&\frac{1}{4\pi\vo}\int_X J\wedge \mc{F}_{\rm vis}\wedge \hat{D}_{\rm vis}
 =\frac{1}{4\pi}\left(q_{{\rm vis}\,4} \, K_4 + q_{{\rm vis}\,5} \, K_5\right), \label{FIvis} \\
 \xi_{\rm int}&=&\frac{1}{4\pi\vo}\int_X J\wedge \mc{F}_{\rm int}\wedge \hat{D}_{\rm int}
 =\frac{1}{4\pi}\left(q_{{\rm int}\,4} \,K_4 + q_{{\rm int}\,5} \, K_5\right), \label{FIint}
\eea
where we denoted $K_i\equiv\partial K/\partial \tau_i$ with $K$ the K\"ahler potential of the 4D effective theory.

\item The chiral intersections between different stacks of D7-branes depend on the gauge fluxes in the following way:
\bea
 I_{\rm vis-E3}&=&\int_X \mc{F}_{\rm vis}\wedge \hat{D}_{\rm vis} \wedge \hat D_{\rm E3}
 =q_{{\rm vis}\,5}, \nonumber \\
 I_{\rm int-E3}&=&\int_X \mc{F}_{\rm int}\wedge \hat{D}_{\rm int} \wedge \hat D_{\rm E3}
 =q_{{\rm int}\,5}, \nonumber \\
 I_{\rm int-vis}&=&\int_X \left(\mc{F}_{\rm vis} - \mc{F}_{\rm int}\right) \wedge \hat D_{\rm vis} \wedge \hat{D}_{\rm int}
 =i_4 q_{{\rm vis}\,4} + i_5 q_{{\rm vis}\,5}-v_4 q_{{\rm int}\,4} - v_5 q_{{\rm int}\,5}. \nonumber
\eea
\end{enumerate}

In order to have an instanton contributing to the superpotential we need to kill its chiral intersections
with the visible sector (and preferably also with any other sector of the theory).
These chiral intersections, if non-vanishing, tend to destroy the instanton contribution,
since they would lead to a superpotential of the form $W \sim \prod_i \Phi_i \, e^{- T_5}$,
where the $\Phi_i$ correspond to open string matter fields,
charged under the visible sector gauge group.
In order to preserve the visible sector group,
these fields must have zero VEVs resulting in a vanishing $W$ \cite{blumen}.
Hence we need to set
$q_{{\rm vis}\,5}=q_{{\rm int}\,5}=0$ which implies $I_{\rm vis-E3}=I_{\rm int-E3}=0$. On the other hand,
we need to have non-vanishing chiral intersections between the D7-stacks on $D_{\rm vis}$ and $D_{\rm int}$.
This is guaranteed by the relations (\ref{chargerelations}) among the moduli charges which ensure that we can
have $q_{{\rm vis}\,4}\neq 0$ and $q_{{\rm int}\,4}\neq 0$ while $q_{{\rm vis}\,5}=q_{{\rm int}\,5}=0$.

Moreover the condition $q_{{\rm vis}\,5}=q_{{\rm int}\,5}=0$ is crucial in order to
keep the modulating field light. In fact, the FI terms (\ref{FIvis}) and (\ref{FIint}),
generically introduce a dependence on both $\tau_5$ and the combination
\be
\tau_4-x\,\tau_5\equiv  \tilde\tau_4\,,
\ee
since
\be
K_4\simeq \frac{3 \alpha \gamma_4 \sqrt{\tilde \tau_4}} {\vo} \qquad \text{and}
 \qquad K_5 \simeq \frac{3 \alpha \left(\gamma_5 \sqrt{\tau_5}- \gamma_4 \,x\sqrt{\tilde\tau_4}\right)} {\vo}.
\ee
Therefore both of these fields would get a very large mass. However,
the condition $q_{{\rm vis}\,5}=q_{{\rm int}\,5}=0$ implies
that $\xi_{\rm vis}$ and $\xi_{\rm int}$ introduce a dependence in the $D$-term potential
{\it only} on the combination $\tilde \tau_4$, leaving $\tau_5$ as a flat direction.
This discussion allows to appreciate the essential r\^ole of the intersection between
the two cycles $D_4$ and $D_5$: thanks to this structure we are left
with a flat direction that we can then use for cosmological purposes.

In order to simplify the system under consideration, we shall also require $q_{{\rm int}\,4}=0$ which implies
$h (\mc{F}_{\rm int})=0$ and $\xi_{\rm int}=0$. The final condition $q_{{\rm vis}\,5}=q_{{\rm int}\,4}=q_{{\rm int}\,5}=0$
written in terms of flux quanta and wrapping numbers reads:
\be
 2 f_4 + v_4 = \frac{ 1 - 2 f_5 - v_5}{x}, \qquad g_4=-\frac{i_4}{2}\,, \qquad g_5=\frac{1-i_5}{2}\,,
\label{conditionSetup}
\ee
which leads to:
\be
q_{{\rm vis}\,4}= 3 {\mathbf b}_4 v_4 \left(2 f_4 + v_4\right)\,, \quad
h (\mc{F}_{\rm vis})= \frac{q_{{\rm vis}\,4}^2}{12 {\mathbf b}_4 v_4}\,, \quad
\xi_{\rm vis}= \frac{q_{{\rm vis}\,4}}{4 \pi}\,K_4\,, \quad I_{\rm int-vis}=  i_4 \,q_{{\rm vis}\,4}\,. \notag
\ee

\subsubsection*{Illustrative flux choice}

An illustrative choice of fluxes that leads to $q_{{\rm vis}\,5}=q_{{\rm int}\,4}=q_{{\rm int}\,5}=0$
and gives rise to three families of chiral matter ({\em i.e.}~$|I_{\rm int-vis}|=3$) is:
\be
f_4=1,\qquad f_5=-63 ,
\qquad g_4=-3,\qquad g_5=0,
\ee
The corresponding
divisors (which we assume to be smooth) wrapped by the visible sector and the intersecting D7-stack are:
\be
v_4=-1\,, \qquad v_5 =126 \qquad \Rightarrow \qquad D_{\rm vis} = 126 D_5- D_4\,,
\ee
and
\be
i_4=6\,, \qquad i_5 =1 \qquad \Rightarrow \qquad D_{\rm int} = 6 D_4+ D_5\,,
\ee
which, from the condition (\ref{conditionSetup}), implies $x=1$ $\Leftrightarrow$ ${\mathbf a}_4={\mathbf a}_5$.
The resulting charge $q_{{\rm vis}\,4}$, flux-dependent shift $h (\mc{F}_{\rm vis})$ and FI-term $\xi_{\rm vis}$ for the visible sector become:
\be
q_{{\rm vis}\,4}= - 3 {\mathbf b}_4\,, \quad
h (\mc{F}_{\rm vis})=-\frac{3 {\mathbf b}_4}{4 }, \qquad \xi_{\rm vis}=-\frac{3 {\mathbf b}_4}{8\pi}\,K_4\,.
\ee

\subsection{Supergravity effective action}

Let us now outline the main features of the effective low-energy
4D supergravity model derived by compactifying over the manifold described in section \ref{compactification}.
After including the leading perturbative corrections, the K\"ahler potential is (we work throughout in the 4D Einstein frame):
\be \label{K0pot}
 K\,\simeq\, K_0+\delta K_{\alpha'}\,+\delta K_{g_s}=\,
 -2\,\ln{\left(\vo+\frac{\xi}{2 g_s^{3/2}}\right)}+\delta K_{g_s}.
\ee
The $\alpha'$ corrections are controlled by the quantity $\xi =-\frac{\zeta(3)\,\chi(X_6)}{2 \,(2 \pi)^3}$,
where $\chi(X_6)$ is the Euler number of the compact manifold $X_6$
while the one-loop open string corrections $\delta K_{g_s}$ take the form studied in \cite{stringloops}.

The non-perturbative superpotential is assumed to have a racetrack dependence on $T_3$,
with additional poly-instantons corrections on $T_5$:\footnote{The non-perturbative corrections dependent on the large cycles $\tau_1$ and $\tau_2$ are negligible and are likely to be absent since these are non-rigid cycles.}
\be \label{W0pot}
 W \simeq W_0 + A_3 e^{- a_3 \left(T_3+ A_5 e^{- 2\pi T_5}\right)} - B_3 e^{- b_3 \left(T_3+ B_5 e^{-2\pi T_5}\right)} \,.
\ee
The superpotential is characterised by the constant $W_0$,
associated with the tree-level flux stabilisation of the dilaton and the complex structure moduli,
and by the non-perturbative corrections weighted by the constants $A_i$ and $B_i$. The parameters $a_3$ and $b_3$ are given by
$a_3 = 2\pi/N_a$ and $b_3=2\pi/N_b$ and arise due to gaugino condensation on D7-branes (with $N_a$ and $N_b$ being the rank of the associated gauge group).
The r\^ole of the racetrack form for the superpotential will become clear in section \ref{potv3p}.

Even though $D_5$ is a rigid cycle ($h^{2,0}(D_5)=0$), single instanton contributions to the superpotential are absent,
due to the presence of extra fermionic zero modes which are Wilson line modulini ($h^{1,0}(D_5)\neq 0$).
However, this zero mode structure can give rise to instanton corrections to the gauge kinetic functions of the
field theories living on $D_3$: this is the origin of the so-called `poly-instanton' corrections which depend on $T_5$.
These tiny non-perturbative effects could be killed by the presence of chiral intersections between the E3 instanton on $D_5$
and visible sector fields living on the cycle $D_4$ that intersects $D_5$.
We have however chosen the gauge fluxes appropriately so to cancel these chiral intersections.
Moreover, this set-up would normally generate $\tau_5$-dependent loop corrections,
which can in principle render this blow-up mode heavy. However, for appropriate choices of fluxes,
the loops depend on the same combination of four-cycles fixed by the $D$-terms, and so
a flat direction remains in the ($\tau_4$, $\tau_5$) plane. We discuss this issue in the next subsection.

\subsection{D-term stabilisation}
\label{dtermapp}

We recall that $D_{\rm vis}$ is the rigid four-cycle supporting a GUT or MSSM-like model in terms of wrapped D7-branes.
It has been shown in \cite{blumen} that the cycle supporting chiral matter cannot get any non-perturbative correction,
since an instanton wrapped around $D_{\rm vis}$ would generically have chiral intersections with visible sector fields.
Therefore the corresponding modulus has to be fixed using different effects \cite{CMV}. This is what we are going to analyse here,
exploiting the field dependence of the $D$-term potential, and of subleading (but crucial) string loop contributions.

The world-volume flux $\mc{F}_{\rm vis}$ generates a modulus-dependent FI-term $\xi_{\rm vis}$.
The resulting $D$-term potential provides the leading order effect that depends on $\tau_4$ and $\tau_5$:
\be
 V_D=\frac {g_{\rm vis}^2}{2}\left(\sum_i c_{{\rm vis}\,i} |\varphi_i|^2-\xi_{\rm vis}\right)^2\,,
\label{Dterms}
\ee
where the gauge coupling $g_{\rm vis}$ is given by $4 \pi g_{\rm vis}^{-2} = v_4\tau_4+v_5\tau_5 - h (F_{\rm vis})  \text{Re}(S)$.
In the expression (\ref{Dterms}) we include also the possible presence of canonically normalised
visible sector singlets $\varphi_i$ (open string states) with corresponding $U(1)$ charges given by $c_{{\rm vis}\,i}$.

We focus on supersymmetric minima where $V_D=0$, for the following reason.
The total scalar potential $V=V_D+V_F$ includes also $F$-term contributions from the matter fields:
\be
 V_F = \sum_i k_i\frac{W_0^2}{\vo^2}|\varphi_i|^2+V_F(T),
\label{FtermPot}
\ee
where the $k_i$ are $\mc{O}(1)$ positive numbers, and $V_F(T)$ denotes the scalar potential for the remaining K\"{a}hler moduli
that we will analyse in detail in the next sections. Given that in a LARGE volume expansion $V_D\sim \vo^{-2}$
whereas $V_F(T) \sim \vo^{-3}$ (as we shall see in the next section),
a non-vanishing $V_D$ would give rise to a dangerous run-away behavior for the volume mode, that must be avoided.

Hence we shall look for non-supersymmetric minima where $V_D$ is
vanishing up to $\mc{O}(\vo^{-1})$ corrections.
The leading order cancellation of the FI-term can be achieved in two ways.
First, the case in which $\xi_{\rm vis}=0$ and $\langle|\varphi_i|\rangle=0$ $\forall i$;
second, the case where $c_{{\rm vis}\,s}\langle|\varphi_s|^2\rangle=\xi_{\rm vis}$ and $\langle| \varphi_i |\rangle=0$ $\forall i\neq s$,
where $\varphi_s$ is a visible sector singlet (like a right-handed sneutrino for example) and $\text{sign}(\xi_{\rm vis})=\text{sign}(c_{{\rm vis}\,s})$.
The first case would fix the combination
$\tilde{\tau}_4 =  \tau_4 - x \tau_5=0$ without forcing any divisor to shrink to zero size.
However, the two-cycle $t_4\propto \sqrt{\tilde{\tau_4}}$ would reach the wall of the K\"{a}hler cone,
{\em i.e.}~$t_4=0$, resulting in a lack of control over the effective field theory.
We need therefore to focus on the second case where a singlet acquires a non-zero VEV.
The minimum for $|\varphi_s|$ is located at (for $\text{sign}(\xi_{\rm vis})=\text{sign}(c_{{\rm vis}\,s})$):
\be
 \langle|\varphi_s|^2\rangle=\frac{\xi_{\rm vis}}{c_{{\rm vis}\,s}}
 \left(1 - \frac{k_s^2 W_0^2}{\lambda\, c_{{\rm vis}\,s}}
 \frac{1}{g_{\rm vis}^2 \sqrt{\tilde\tau_4} \vo}\right)\,,\qquad\text{where}
 \qquad \lambda\equiv \frac{3 \alpha \gamma_4 q_{{\rm vis}\,4}k_s}{4\pi c_{{\rm vis}\,s}}>0\,.
\ee
Substituting this value in the total scalar potential $V=V_D+V_F$ we find:
\be
V= \lambda W_0^2 \frac{\sqrt{\tilde\tau_4}}{\vo^3}-\mu\frac{v_4\tau_4+v_5\tau_5}{ \vo^4}+V_F(T)\,,
\qquad\text{where}\qquad\mu\equiv \frac{k_s^2 W_0^4}{8\pi c_{{\rm vis}\,s}^2 }\,.
\label{NewVtotale}
\ee
As we shall see in the next section, the Hubble constant during inflation scales
as $H^2\sim W_0^2 \,\vo^{-10/3}$ (setting $M_P =1$), whereas the mass-squared of the canonically normalised modulating field $\sigma$
scales as $m_{\sigma}^2\sim \vo \,V$. This implies that in order to have a light modulating field,
the potential for this field has to be developed at order $V<\mc{O}\left(W_0^2\vo^{-(4+1/3)}\right)$.
However the potential (\ref{NewVtotale}) introduces a dependence on the combination $v_4 \tau_4 + v_5 \tau_5$
at order $\mc{O}\left(W_0^4 \vo^{-4}\right)$ which would render the modulating field too heavy if $W_0\sim \mc{O}(1)$.
Hence we have to perform a very moderate tuning of $W_0$ of the order $W_0< \vo^{-1/6}$. \footnote{In the next section,
we shall fix this leading order flat direction via a poly-instanton generated potential which scales as
$V_{\rm poly}\sim \mc{O}\left(W_0^2\vo^{-(3+p)}\right)$ and so we shall, more correctly, tune $W_0$ such that
$W_0 \lesssim \vo^{(1-p)/2}$. Moreover, the modulating field could be kept light also by choosing $v_4$ and $v_5$
such that the combination $v_4 \tau_4 + v_5 \tau_5 =\tilde\tau_4$ ({\em i.e.}~$v_4=1$ and $v_5=-x$). However we do not consider this option since
it would lead to a modulation mechanism which would not be able to beat the standard mechanism for the production of the density perturbations.}

After having analysed $D$-term and matter field $F$-term contributions to the potential for $\tilde \tau_4$, we examine
how it is influenced by string loop corrections, showing that the latter are able to fix this combination.
Given that each cycle wrapped by a stack of D7-branes receives
one-loop open string corrections \cite{GenAnalofLVS, stringloops},
one would expect effects of order $\mc{O}(\vo^{-3})$ which depend on both $\tau_4$ and $\tau_5$.
However, the coefficients of these $g_s$ corrections can be appropriately tuned
so that at leading order they depend only on the
combination $\tilde{\tau}_4$, whereas all the other terms are very suppressed.
The leading order loop correction then reads \cite{stringloops}:
\be
 V_{g_s}= \frac{C_{\rm loop} W_0^2}{\vo^3 \sqrt{\tilde{\tau_4}}}\,.
 \label{V at 1-loop}
\ee
These loop effects can indeed compete with the $\tilde{\tau_4}$-dependent $F$-term piece in (\ref{NewVtotale}). The complete scalar potential for the combination $\tilde{\tau_4}$ is then constituted by the sum of
all contributions:
\be
 V (\tilde{\tau_4}) = \frac{W_0^2}{\vo^3}\left(\lambda \sqrt{\tilde{\tau_4}}+\frac{C_{\rm loop}}{\sqrt{\tilde{\tau_4}}}\right)\,,
\label{VforTilde}
\ee
and it admits a minimum at $\langle \tilde{\tau_4}\rangle=C_{\rm loop}/\lambda$.
If we integrate out $\tilde{\tau_4}$ and place it to its minimum we are left with a volume dependent potential of the form:
\be\label{voldiptau4}
 V (\tilde{\tau_4}=\langle\tilde{\tau_4}\rangle) = 2 \sqrt{\lambda C_{\rm loop}}\,\frac{W_0^2}{\vo^3}\,,
\ee
that is proportional to $1/\V^3$.
This shows that, as anticipated, $D$-term, $F$-term, and string loop contributions to the moduli potential
stabilise the linear combination $\tilde \tau_4$, leaving a leading order flat direction
which will be softly lifted by polyinstanton corrections, and will play the r\^ole or our modulating field.

\subsection{F-term stabilisation}
\label{Fstab}

We now study the structure of the $F$-term scalar potential for the K\"ahler moduli using the K\"ahler potential and the superpotential of eqs.~\pref{K0pot} and \pref{W0pot}. We follow  the discussion of ref.~\cite{fiberinfl} and \cite{StringyADD}.
The scalar potential can be organised in an expansion in inverse  powers of the volume:
\be
 V=V_{\mc{O}(\vo^{-3})}+V_{\mc{O}(\vo^{-10/3})}
 +V_{\mc{O}(\vo^{-3-p})},\text{ \ \ with \ \ }p>1/3.
\ee
where $p$ is a number that depends on the parameters of the model (more on this later).  We now discuss the previous potential order by order in the expansion, showing that at each order it can provide stable minima for some of the moduli.

\subsubsection*{Moduli stabilisation at $\mc{O}(\vo^{-3})$}

After minimising the axionic $\tau_3$ direction, as usual in models based on LVS, and trading $\tau_2$ for the volume $\vo$ (in the limit in which $\sqrt{\tau_1} \tau_2 \gg \tau_i^{3/2}$ for $i>2$), the leading order contribution to the $F$-term scalar potential can be readily calculated and results:
\footnote{Since $\gamma_3$ does not play any relevant r\^ole in the discussion, from now on we use it to set $\alpha  \gamma_3  = 1$ to simplify some of the formulae. Moreover, we  neglect the overall normalisation factor $g_s e^{K_{cs}}/(8\pi)$ \cite{BCGQTZ}.}
\begin{eqnarray}
 V_{\mathcal{O}(\vo^{-3})}&=&\frac{8 \sqrt{\tau_3} \left(A_3^2 a_3^2 e^{-2 a_3 \tau_3}-2 A_3 B_3 a_3 b_3 e^{- a_3 \tau_3-b_3 \tau_3}
 +B_3^2 b_3^2 e^{-2 b_3 \tau_3}\right)}{3 \vo} \notag \\
 &&+\frac{4 W_0 \tau_3  \left(A_3 a_3 e^{-a_3\tau_3}-B_3 b_3 e^{-b_3\tau_3}\right)}{\vo^2}+\frac{3 W_0^2 \xi }{4 g_s^{3/2}\vo^3}.
\label{LeadingV}
\end{eqnarray}
The scalar potential, at this order in the expansion, contains  terms with different signs associated with the racetrack structure of the superpotential. Eq.~(\ref{LeadingV}) does not depend on the fibre modulus $\tau_1$ \cite{fiberinfl}, nor  on the blow-up mode $\tau_5$ supporting the poly-instanton corrections \cite{StringyADD}. It does not do so because the dominant contribution to the potential of $\tau_1$ arises via string loops \cite{GenAnalofLVS} at order $\mc{O}(\vo^{-10/3})$. As we show later, $\tau_5$ develops a potential at even more subleading order, $\mc{O}(\vo^{-3-p})$, where $p$ is a parameter that assumes values  larger than $1/3$.

On the other hand, as we learned in the previous section, combined effects of $D$-term, $F$-term,  and string loop contributions do stabilise the combination $\tilde \tau_4$: upon integrating it out, we are left with a volume dependent contribution to the potential summarised in eq. (\ref{voldiptau4}), which can be simply regarded  as a shift of the $\alpha'$ correction term:
\be\label{stabv3}
 \frac{3 W_0^2 \xi }{4 g_s^{3/2}\vo^3} \,\,\,\rightarrow\,\,\, \frac{3 W_0^2 \Xi }{4 g_s^{3/2}\vo^3},
 \text{ \ \ where \ \ }
 \Xi=\xi + \frac{8}{3} \,g_s^{3/2}\sqrt{\lambda C_{\rm loop}}\,.
\ee
We now focus on the potential (\ref{LeadingV}), replacing $\xi$ with $\Xi$.
In order to express the conditions for the stabilisation of $\tau_3$ and $\vo$ in a compact way,
it is convenient to introduce the quantity $s_3\equiv a_3-b_3$.
The vanishing of the first derivatives of $V_{\mathcal{O}(\vo^{-3})}$
with respect to $\tau_3$ and $\vo$ provides the following extrema for $\tau_3$ and $\vo$
(keeping the first order contributions in an expansion in $1/(a_3\tau_3)$ and $1/(b_3\tau_3)$):
\be
 \langle \tau_3 \rangle=\,\left(
 \frac{\Xi}{2} \right)^\frac23\frac{1}{g_s}\,,
 \text{ \ \ \ \ }e^{- b_3 \langle \tau_3\rangle }
 =\frac{3 W_0 \sqrt{\langle \tau_3\rangle }}{4 Z \langle \vo\rangle}f_1\,,
\label{mint3}
\ee
where
\be
 Z\equiv B_3 b_3 -A_3 (b_3+s_3) e^{-s_3 \langle \tau_3\rangle}\,,
\ee
and
\be
 f_1\equiv 1-\,\frac{3\,f_2}
 {1 + s_3 \left( \frac{1}{b_3}-\frac{B_3}{Z}\right)}\,,
\label{MinTau3}
\ee
with $f_2\equiv 1/(4 b_3 \langle\tau_3\rangle)\ll 1$ for large $b_3\langle\tau_3\rangle$. We also impose that $Z$ is  positive in order to have a minimum in the $\tau_3$-axion direction. The extrema associated with equation (\ref{mint3}) correspond to  stable minima both for the volume  $\V$ (or alternatively the field $\tau_2$) and the field $\tau_3$. These formulae show that the volume is LARGE thanks to an exponential dependence on the parameters of the model.

Notice that in the case  of single exponential ($s_3=0$), (\ref{MinTau3}) reduces to $f_{1}=1-3 f_2$, implying that for $b_3\langle\tau_3\rangle\simeq \ln \vo\gg 1$ (the limit of large volume we are interested in) the corrections to $f_1$ due to $f_2$ are always subleading. As we will discuss in what follows, the consistency of our approximations requires to avoid this, and so we set $s_3\neq 0$. We finally point out that this minimum is AdS and so additional uplifting terms are needed to uplift the potential to a nearly Minkowski vacuum (for explicit de Sitter examples see \cite{Up}).

\subsubsection*{Moduli stabilisation at order $\mc{O}(\vo^{-10/3})$}\label{potv103}

Each cycle wrapped by a stack of D7-branes receives one-loop open string corrections \cite{GenAnalofLVS, stringloops} which, as pointed out in \cite{fiberinfl}, generate a subleading potential for $\tau_1$. This potential, for a certain range of $\tau_1$,   results flat enough to be suitable for inflation. In this subsection we focus only on the $\tau_1$ and $\tau_2$-dependent
loop corrections which can be estimated using a procedure identical to \cite{fiberinfl}:
\be
 V_{\mc{O}(\vo^{-10/3})}=\left(\frac{A}{\tau_1^2}
 -\frac{B}{\vo\sqrt{\tau_1}} +\frac{C\tau_1}
 {\vo^2}\right)\frac{W_0^2}{\vo^2},
 \label{potwl}
\ee
where $A$, $B$, $C$ are given by:
\bea
 A&=&\left( g_s C_1^{KK}\right)^2>0,\text{ \ \ \ }B=4 C_{12}^{\,W},\text{ \ \ \ }
 C= 2\left( g_s\,C_2^{KK}\right)^2>0 \,,
 \label{defC}
\eea
where $C_1^{KK}$, $C_{12}^{\,W}$, and $C_2^{KK}$ are constants
that depend on the VEV of the complex structure moduli which have been fixed at tree level (see \cite{fiberinfl} for more details). In what follows we regard these constants as free,  to be fixed using consistency or phenomenological
requirements. When $32 A C\ll B^2$ $\Leftrightarrow$ $4\, g_s^4 \left(C_1^{KK} C_{12}^W\right)^2\ll \left(C_2^{KK}\right)^2$,
the minimum for $\tau_1$ is at:
\be
  \langle \tau_1 \rangle
 \simeq \left(-\frac{B \langle\vo\rangle}{2C} \right)^{2/3}
  \quad \hbox{if $B<0$}
  \qquad \hbox{or} \qquad
 \langle \tau_1 \rangle \simeq
 \left(\frac{4A \langle\vo\rangle}{B} \right)^{2/3}
  \quad \hbox{if $B>0$} \,, \label{tau1soln2}
\ee
justifying the $\vo^{-10/3}$ scaling of (\ref{potwl}). In the following, for definiteness, we consider the case $B>0$.

\subsubsection*{Moduli stabilisation at $\mc{O}(\vo^{-3-p})$}
\label{potv3p}

The leading order contribution to the scalar potential from the $\tau_5$-dependent poly-instanton corrections
turns out to be more suppressed than the ones analysed in the previous subsection since it scales as \cite{StringyADD}:
\begin{eqnarray}
 V_{\mathcal{O}(\vo^{-3-p})} &=&\frac{16\sqrt{\tau_3}e^{-2\pi\tau_5}}{3\vo}\left[
 a_3^3
 A_3^2  A_5 e^{-2 a_3 \tau_3}
 +  b_3^3  B_3^2 B_5 e^{-2 b_3 \tau_3}- a_3
 b_3  A_3 B_3 e^{-a_3\tau_3-b_3\tau_3}\left(a_3 A_5 +b_3 B_5\right)\right] \notag \\
 &&  -\frac{4 W_0 e^{-2\pi\tau_5}}{\vo^2}\left[B_3 b_3 B_5 \left(b_3\tau_3+2\pi\tau_5\right) e^{- b_3 \tau_3}
 - A_3 a_3 A_5 \left(a_3\tau_3+2\pi\tau_5\right) e^{-a_3 \tau_3} \right], \label{Vp}
\end{eqnarray}
where we have neglected $\tau_5$-independent pieces in $V_{\mathcal{O}(\vo^{-3-p})}$, since they are not relevant for moduli stabilisation.
Substituting (\ref{mint3}) in (\ref{Vp}), we obtain a very compact expression
\be
 V_{\mathcal{O}(\vo^{-3-p})}=-\frac{A_0 W_0^2}{\vo^3}
 \left(2\pi \tau_5-p \,b_3\langle \tau_3\rangle\right) e^{-2\pi \tau_5},
\label{V}
\ee
where
\begin{eqnarray}
 A_0&=&3 \sqrt{\langle \tau_3 \rangle} \,\frac{\left( A_5 Z+S_5 B_3 b_3\right)}{Z}f_{1}\sim \mathcal{O}(1),
 \qquad\text{with}\qquad S_5\equiv B_5-A_5\notag \\
 p&=&\left[\frac{s_3
 B_3(A_5+S_5)}{A_5 Z+S_5 B_3 b_3}-\frac{(b_3+s_3)}{b_3}\right](1-f_{1}).
\end{eqnarray}
The potential (\ref{V}) for $A_0>0$ admits a minimum at
$2\pi \langle \tau_5 \rangle=p\, b_3 \langle\tau_3\rangle+1\simeq p\, b_3 \langle \tau_3 \rangle$
and so it scales as $\vo^{-3-p}$ (the value of $A_0$ determines the depth of the vacuum).

At this point, we can finally fully appreciate the r\^ole of the racetrack form of the superpotential.
Let us consider the contributions proportional to $f_2$ to the parameter $p$
\be
 p \,=-
 \left[1+s_3 A_5\left(\frac{Z-B_3 b_3}{b_3(A_5 Z+S_5 B_3 b_3)}\right)\right]
 \,
 \left(\frac{3\,f_2}
 {1+s_3 \left(\frac{1}{b_3}-\frac{B_3}{Z}\right)}\right).
\ee
In the single exponential case (setting $s_3=0$), $p$ becomes negative with an absolute value smaller than unity:
\be
 |p|=3 f_2 =\frac{3}{4 b_3\langle\tau_3\rangle}\ll 1\,\,\, \Rightarrow \,\,\,
 2\pi \langle\tau_5\rangle = -\frac{3}{4}<0,
\ee
and so we end up in a regime where the minimum for $\tau_5$ is out of the K\"ahler cone.
However in the racetrack case it is possible to render $p$ positive and large enough to trust the effective field theory.
As we will learn in section \ref{sec-reheating}, a value of $p$ in the interval $4/3 < p <22/9$ is preferred for our cosmological applications.
Many possibilities exist, although the choice of the
parameters have to be carefully balanced in order to get the desired result.
Focusing on the illustrative flux choice of section \ref{set-up}, we present here a parameter fit that provides an acceptable
minimum for $\tau_5$, and a value of $p$ in the preferred range (recall that  $a_3=2\pi/N_a$ and $b_3=2\pi/N_b$):
\bea\label{secfit}
 k_{122}&=&1,\quad C_{\rm loop}=35.17,\quad k_s=1,\quad c_{{\rm vis}\,s}=-1, \quad
 \xi\simeq 5.308,\nn \\ g_s&\simeq&0.2058,\quad W_0=0.1,\quad A_3=1, \quad B_3=15061.5,\quad N_a=4, \quad N_b=3, \nn \\
 \quad C_1^{KK}&\simeq&15.26,\quad C_{12}^W=0.01,\quad A_5=-124.445,\quad B_5=-1.
\eea
This choice of parameters yields the following value for the VEVs of the fields at the minimum:
\bea
 \langle\tau_1\rangle
 &\simeq& 28965.8,\quad
 \langle\vo\rangle\simeq 5000,
  \quad
 \langle\tau_3\rangle \simeq 9.72,\text{ \ \ }\langle\tilde\tau_4\rangle\,=\, 625,
 \quad
 \langle\tau_5\rangle\simeq 5, \nn \\
 && \Rightarrow \langle\tau_2\rangle\simeq 41.55,\quad \langle\tau_4\rangle \simeq 630, \quad p\simeq\,1.5 \,.
\eea
Notice that $p$ lies in the interval preferred by our cosmological applications. Since it is larger than $1/3$,
the resulting potential for the field $\tau_5$ is suppressed by higher powers of the volume with respect to the string-loop potential.
Moreover our choice of $W_0$ guarantees that the term in (\ref{NewVtotale}) proportional to $\mu$
which arises from the non-exact cancellation of the $D$-term potential, does not beat the poly-instanton potential.
In fact, this requirement turns into the condition $W_0 \lesssim \vo^{(1-p)/2}\simeq 0.12$.

This fit determines and fixes most of the available parameters in this set-up, and has also 
interesting cosmological consequences as we will discuss in section \ref{sect:nongau}.
The only exceptions are parameters characterising the geometry $x$ and ${\mathbf b}_4$,
and the parameter $C_2^{KK}$ characterising the string loop potential.
In what follows, we will fix the quantities as in the previous fit
and use the remaining undetermined parameters
to further characterise the predictions for the cosmological model of interest.

\subsection{Kinetic terms and canonical normalisation}
\label{sect-can}

After determining an expression for the potential,
we need to analyse the kinetic terms in order to canonically normalise the fields.
As for the potential, also the kinetic Lagrangian
can be organised in terms of an expansion in inverse powers of the volume:
\be
 {\cal L}_{kin} = {\cal L}_{kin}^{\mathcal{O}(1)}+{\cal
 L}_{kin}^{\mathcal{O}(\vo^{-1})}\,. \label{LagKin}
\ee
We analyse in detail the procedure to render canonical
the kinetic terms in appendix \ref{app-kin}.
There we show how to define new canonically normalised fields,
dubbed $\phi$, $\sigma$, $\chi_2$, $\chi_3$, $\chi_4$.
In terms of these fields, the kinetic Lagrangian,
up to second order in inverse powers of the volume, reads simply:
\be
 {\cal L}_{kin} = \frac{1}{2} \,\partial \phi^2+\frac{1}{2} \,\partial \sigma^2+
 \frac12 \sum_{i=2}^4 \partial \chi_i^2.
 \ee
Here we only present the results,
and write the relations between the original fields $\tau_i$
and the canonically normalised fields (trading $\tau_2$ for $\V$):
\bea
 \tau_1 &=& \exp{\left(\frac{2}{\sqrt{3}}
 \, \phi + \sqrt{\frac23} \, \chi_2
 + \frac{1}{2}\sum_{j=3}^4 \chi_j^2 +\frac12 \sigma^2\right)}, \label{TAU1}
\\
 \vo &=& \exp{\left(\sqrt{\frac32} \, \chi_2+ \kappa\left(\phi,\chi_2\right)\phi \right)} \,, \label{Vol}
\\
 \tau _{3}\, &=&\,\left( \frac{3}{4\alpha
 \gamma_{3}}\right) ^{\frac{2}{3}}
 \exp {\left(\sqrt{\frac23} \,\chi_2+ \frac{2}{3}\kappa\left(\phi,\chi_2\right)\phi   \right) }
 \,\chi_{3}^{\frac{ 4}{3}}\,,
 \label{Smalltaus}
\\
  \tau_4&=&\left( \frac{3}{4\alpha}\right)^{\frac{2}{3}}
 \exp {\left(\sqrt{\frac23} \,\chi_2+ \frac{2}{3}\kappa\left(\phi,\chi_2\right)\phi\right) }
 \left(\frac{\chi_4^{4/3}}{\gamma_4^{2/3}}+x\frac{\sigma^{4/3}}{\gamma_5^{2/3}}\right)\,,
 \label{tau4canonica}
\\
 \tau_5&=&\left( \frac{3}{4\,\alpha \gamma_5}\right) ^{\frac{2}{3}}
 \exp {\left(\sqrt{\frac23} \,\chi_2+ \frac{2}{3}\kappa\left(\phi,
 \chi_2\right)\phi\right)} \,\sigma^{\frac{ 4}{3}}\,,
 \label{tau5canonica}
\eea
where the function $\kappa\left(\phi,\chi_2\right)\sim \mc{O}\left(\vo^{-1/3}\right)$
is a subleading correction which in the limit of exponentially large volume and large $\tau_1$ takes the form:
\be
 \kappa\left(\phi,\chi_2\right)=\left(\frac{2}{3}\right)^{15/4}\,
 \sqrt{\frac{2 \,b_3^3}{\chi_2}}\,
 e^{-\frac{1}{\sqrt{6}}\chi_2}
 \left(B \,e^{-\frac{1}{\sqrt{3}}\,\phi}-4 A\,e^{-\frac{4}{\sqrt{3}}\,\phi}\right).
\ee
Notice that the combination $\tilde\tau_4$ can be written in terms of the canonically normalised fields as
\be
 \tilde\tau_4=\left( \frac{3}{4\alpha\gamma_4}\right)^{\frac{2}{3}}
 \exp {\left(\sqrt{\frac23} \,\chi_2+ \frac{2}{3}\kappa\left(\phi,\chi_2\right)\phi\right)}\chi_4^{4/3}\,.
 \label{tilde4}
\ee
whereas the leading order volume scaling of the minimum of the canonically normalised fields is:
\bea
 \langle \phi \rangle&\simeq& \frac{\sqrt{3}}{2}\,\ln \langle\tau_1 \rangle-\frac{1}{\sqrt{3}}\,\ln\vo,\qquad
 \langle \chi_2 \rangle\,\simeq\,
 \sqrt{\frac23}
 \,\ln{\V},\qquad
 \langle \chi_3 \rangle\,\simeq\,\sqrt{\frac{4\,\alpha \gamma_3}{3}} \,\langle \tau_3 \rangle^{\frac34} \,\V^{-\frac12},
 \nonumber\\
 \langle \chi_4 \rangle&\simeq& \sqrt{\frac{4\,\alpha \gamma_4}{3}}\, \langle  \tilde\tau_4 \rangle^\frac34 \,\V^{-\frac12},
 \qquad
 \langle \sigma \rangle\,\simeq\,
 \sqrt{\frac{4\,\alpha \gamma_5}{3}}\,
 \langle \tau_5 \rangle^\frac34 \,\V^{-\frac12},
\label{vevsnf}
\eea
where the values of the original fields at their minimum, as obtained in the previous subsections, are given by (at leading order in the volume):
\be
 \langle \tau_1 \rangle\simeq\left(
 \frac{4 A\, \V}{B}
 \right)^{\frac23},\qquad  \langle \tau_3 \rangle\,\simeq\,
 \,\frac{\ln{\V}}{b_3}, \qquad \langle\tilde\tau_4\rangle \simeq\frac{C_{\rm loop}}{\lambda}\,,
 \qquad 2\pi \langle \tau_5 \rangle\,\simeq\,
 p\, b_3 \langle \tau_3 \rangle\,.
\ee

\section{Inflationary and post-inflationary dynamics}
\label{sec-infandpinfdyn}

After having acquainted a good theoretical control on the scalar potential and kinetic terms, we turn to study how inflation is realised in this set-up.

\subsection{Moduli dynamics during inflation}
\label{sec-infl}

We assume that inflation occurs while one of the moduli we consider is rolling over the potential to reach its minimum.
In our set-up, we have two light moduli that can have interesting cosmological applications: $\phi$ and $\sigma$. For our purposes,
as we discuss in section \ref{sect-gendis}, the best candidate for driving the inflationary process is the field $\phi$, while the field $\sigma$ modulates the reheating process.\footnote{The possibility
to drive inflation using the field $\sigma$ controlled by poly-instanton corrections has been recently studied in
\cite{PolyInfl}.}
We can assume that the  fields $\chi_2$, $\chi_3$ and $\chi_4$ lie at their minima during inflation: we check later that these fields have indeed masses much larger than the Hubble scale during inflation, so this hypothesis is correct.

We start discussing the inflationary dynamics. Since the potential for the inflationary field $\phi$ is controlled by string loop contributions, the inflationary dynamics is very similar to the one of \emph{fibre inflation} \cite{fiberinfl}. Expanding $\tau_1$ in terms of its canonically normalised counterpart (and writing $\varphi = \langle \varphi \rangle + \hat \varphi $ to denote the classical shift of some field $\varphi$ from its minimum) allows to write, at leading order in inverse powers of the volume:
\be
 \tau_1\simeq\langle\tau_1\rangle \,e^{\frac{2}{\sqrt3}\,\hat\phi} =
 \left(q \vo\right)^{2/3}\,e^{\frac{2}{\sqrt3}\,\hat\phi}, \quad
 \text{ \ where \ } \quad q\equiv 4 A/B.
\ee
The potential for the field $\phi$ is provided by string loops corrections, as we discussed in section \ref{potv103}. It is given by eq. (\ref{potwl}),  expressing $\tau_1$ in terms of the canonically normalised field $\hat \phi$:\footnote{We introduced the overall normalisation factor $g_s/8\pi$ (see \cite{BCGQTZ}). Unless otherwise stated, we set the Planck mass $M_P=1$.}
\be
 V_{\rm inf}(\hat\phi) =\frac{g_s\,W_0^2 \,A \, q^{4/3}}{8\pi
 \,\langle\vo\rangle^{10/3}}\,\left[
 \frac{c}{2}\,e^{\frac{2}{\sqrt{3}} \,\hat \phi
}-4 \,e^{-\frac{1}{\sqrt{3}}\, \hat \phi }
 +  e^{-\frac{4}{\sqrt{3}}\,  \hat \phi}+3 - \frac{c}{2}
 \right] \,,
\label{Inflpot}
\ee
with (see eq.~(\ref{tau1soln2})) $c \equiv 32 A\, C/B^2 \simeq 4 \cdot 10^{-5} \ll 1$ for $C_2^{KK}=4.8 \cdot 10^{-5}$.
Most of the parameters appearing in the previous potential are fixed by the fit in eq. (\ref{secfit}),
that is by the requirement of having controlled minima for the stabilised moduli.

\begin{figure}
\begin{center}
\scalebox{.6}{\includegraphics*{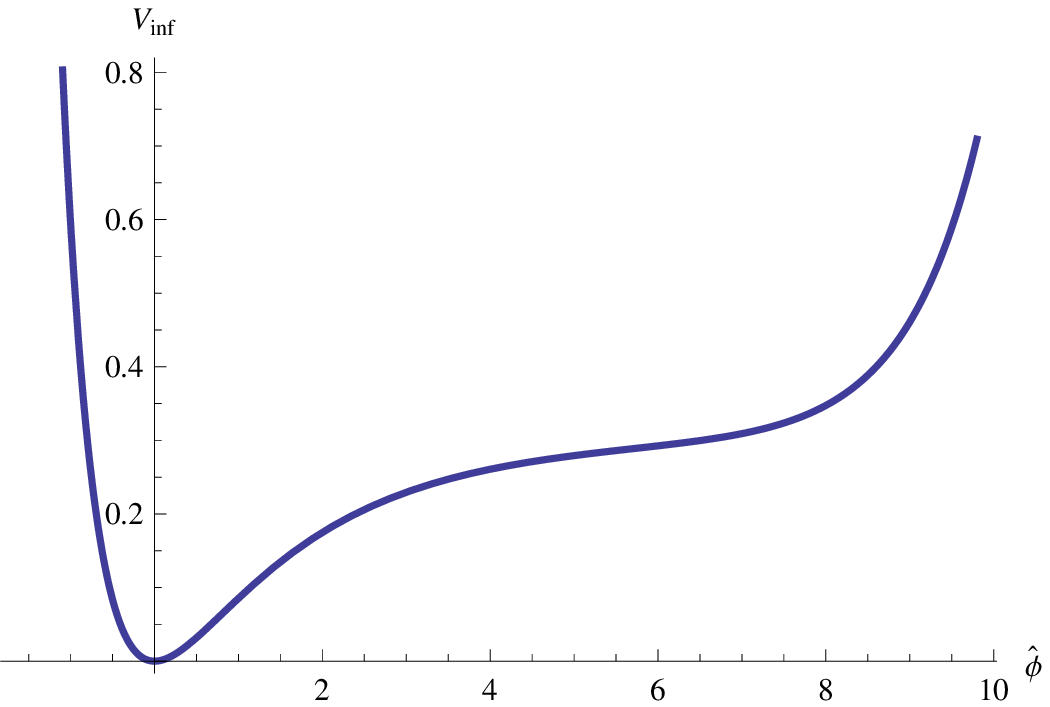}}
$\qquad\qquad$
\scalebox{.6}{\includegraphics*{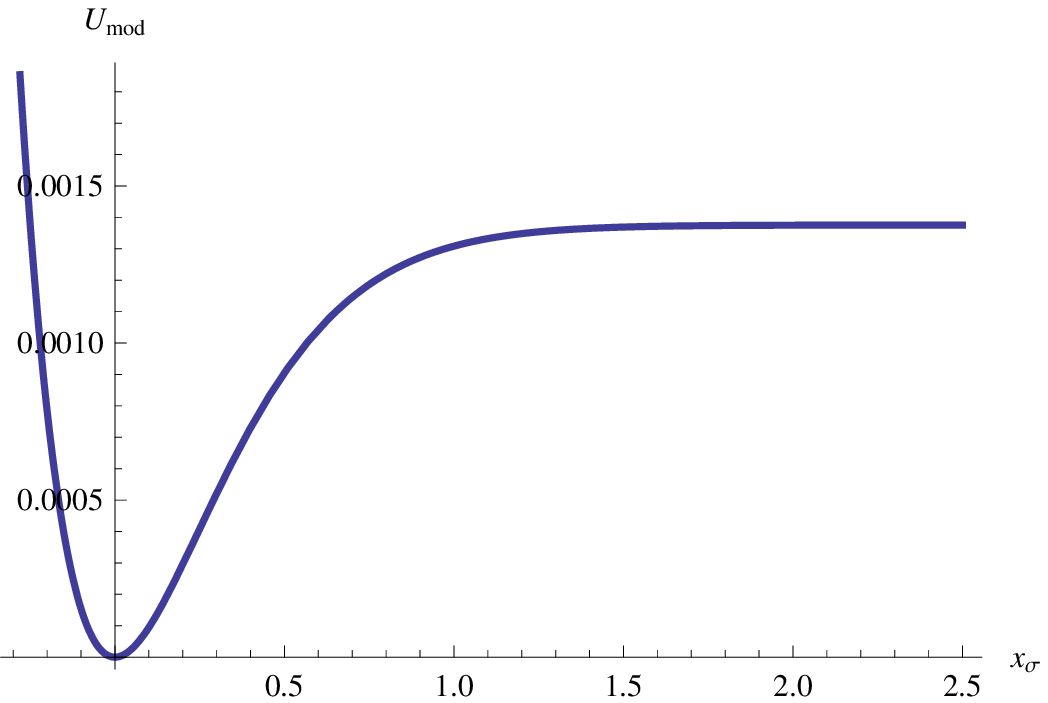}}
\caption{The potentials for the inflaton (left) and the modulating field (right). Inflation occurs in the flat region
of the inflaton potential while the modulating field gets large quantum oscillations around its classical value
which is almost frozen during inflation at a value which is not necessarily close to its minimum.}\label{fig1}
 \end{center}
\end{figure}

The slow roll parameters derived from the inflationary potential (\ref{Inflpot}) take the form \cite{fiberinfl}:
\bea
 && \epsilon \equiv \frac{1}{2} \left(\frac{V_{\rm inf}'}{V_{\rm inf}}\right)^2 =
 \frac{1}{6}\frac{\left[c \,e^{\frac{2}{\sqrt{3}} \,\hat \phi}
 + 4 \,e^{-\frac{1}{\sqrt{3}}\, \hat \phi }
 -4  \,e^{-\frac{4}{\sqrt{3}}\,  \hat \phi} \right]^2}{\left[\frac{c}{2}\,e^{\frac{2}{\sqrt{3}} \,\hat \phi
 }-4 \,e^{-\frac{1}{\sqrt{3}}\, \hat \phi }
 +  e^{-\frac{4}{\sqrt{3}}\,  \hat \phi}+3 - \frac{c}{2}\right]^2} \,,  \\
 && \eta \equiv \frac{V_{\rm inf}''}{V_{\rm inf}} =
 \frac{1}{3}\frac{\left[2 c \,e^{\frac{2}{\sqrt{3}} \,\hat \phi}
 - 4 \,e^{-\frac{1}{\sqrt{3}}\, \hat \phi }
 +16  \,e^{-\frac{4}{\sqrt{3}}\,  \hat \phi} \right]^2}{\left[\frac{c}{2}\,e^{\frac{2}{\sqrt{3}}\, \hat \phi
 }-4 \,e^{-\frac{1}{\sqrt{3}}\, \hat \phi }
 +  e^{-\frac{4}{\sqrt{3}}\,  \hat \phi}+3 - \frac{c}{2}\right]^2} \,.
\eea
Inflation occurs when $\epsilon\ll 1$ and $\eta \ll 1$,
which happens in a region where the potential $V_{\rm inf}$ is sufficiently flat (see Fig.\ref{fig1}).
From the expressions above, one notices that a sufficient condition to ensure this is
$c\ll 4 \,e^{-\sqrt{3}\, \hat\phi}$ and $e^{-1/\sqrt{3}\, \hat\phi}\ll1$. The number of e-foldings is given by:
\be
 N_e= \int_{\hat\phi_{\rm end}}^{\hat\phi_*} \frac{V_{\rm inf}}{V_{\rm inf}'} \,
 d\hat\phi\simeq
 \frac{\sqrt{3}}{4} \int_{\hat\phi_{\rm end}}^{\hat\phi_*} \left[\left(3-\frac{c}{2}\right)
 e^{\hat\phi/\sqrt{3}} -4 \right] d\hat\phi = \frac{3}{4}
 \left[\left(3-\frac{c}{2}\right) e^{\hat\phi/\sqrt{3}} -\frac{4 \hat \phi}{\sqrt{3}}\right]_{\hat\phi_{\rm end}}^{\hat\phi_*}, \nonumber
\ee
where starred  quantities are evaluated at horizon exit. During the inflationary period,
the Hubble parameter is given by the following expression:
\be
 H^2\,\simeq\,\frac{g_s \, W_0^2\,A\,q^{4/3}}{8\pi\,\langle\vo\rangle^{10/3}}\simeq 3.7 \cdot 10^{-12}\,.
\ee
Notice that it scales with the volume as $H^2\sim  1/\V^{10/3}$, a reference scale for the remaining discussion.

Armed with these preliminary definitions, and using the parameter fit of eq. (\ref{secfit}),
we can determine an inflationary trajectory characterised by the following features:
\be
\phi_{\rm end}=1 \quad \text{where}\quad \epsilon_{\rm end}\simeq 0.78,\qquad \text{and}\qquad
\phi_*=6.026 \quad \text{where}\quad \epsilon_*\simeq 5.4\cdot 10^{-4},
\ee
which yield:
\be\label{inflprop}
 N_e \simeq 60 \,, \qquad\text{and}\qquad\left(\frac{{\mathcal P}_{\zeta_\phi}}{ {\mathcal
 P}_{\zeta\,{\rm COBE}}}\right)^{1/2}
  \simeq 0.018\,.
\ee
Notice that once $C_2^{KK}$ has been fixed, all the quantities characterising the inflationary process are determined.
We get a reduced amplitude of the scalar power spectrum, defined as:
\be\label{inflps}
 {\mathcal
 P}_{\zeta_\phi}
 \,=\,H^2/(8 \pi^2 \epsilon)\,,
\ee
with respect to the COBE normalisation $ {\mathcal P}_{\zeta\,{\rm COBE}} =  2.7 \times 10^{-7}$. This is not a problem since the curvature perturbations will be produced by the alternative mechanism of modulated reheating, as we are going to discuss in section \ref{sec-modreh}. Notice that the value of the parameter $\epsilon$ is relatively large in this set-up, as in \emph{fibre inflation} \cite{fiberinfl}.
In the case in which the inflaton provides the dominant contribution to the curvature fluctuations, this would imply that the tensor-to-scalar ratio $r_T$ is large (this was indeed one of the motivations to study \emph{fibre inflation} \cite{fiberinfl}); in our context, this is not so since the curvature perturbations are produced by a different mechanism. Nevertheless, having a large $\epsilon$ will be important for having a successful scenario of modulated reheating.

During inflation, the masses of the $\chi$-fields, once
canonically normalised, scale with the volume as:\footnote{The volume scalings of the various fields can be understood as follows:
from (\ref{vol}), at leading order $\chi_2$ is mostly
the overall volume which is stabilised at order
$\mc{O}(\vo^{-3})$. Both $\tau_3$ (mostly given by $\chi_3$ as can be seen from (\ref{smalltaus})) and $\tilde{\tau_4}$
(mostly given by $\chi_4$ as can be seen still from (\ref{smalltaus})) are fixed at order $\mc{O}(\vo^{-3})$. The potential for  $\sigma$ is discussed next.}
\begin{eqnarray}
 m_{\chi_2}^2&\sim&
 \frac{W_0^2}{\vo^{3}}, \text{ \ \ \ \ \ }
 m_{\chi_3}^2 \sim\frac{W_0^2}{\vo^2}\,, \label{seconda}
 \text{ \ \ \ \ }
 m_{\chi_4}^2 \sim \frac{W_0^2}{\vo^2}\,.
\end{eqnarray}
For our  large value of the volume, $\chi_2$, $\chi_3$ and $\chi_4$  have masses larger than the Hubble parameter $H^2 \sim W_0^2/\V^{\frac{10}{3}} $ during inflation: this shows that our hypothesis that they sit on their minima is consistent. The field $\sigma$ can instead be light during inflation, for our value of  $p = 1.5$. The potential for $\sigma$ is calculated starting from the potential for the field $\tau_5$, analysed in section \ref{potv3p}. Expanding the modulating field as $\sigma = \langle\sigma\rangle+\hat\sigma$, the field $\tau_5$ is given in terms of its canonically normalised counterpart by:
\be
 \tau_5 =\langle\tau_5\rangle\left(1+x_{\sigma}\right)^{4/3}\,,
 \qquad\text{where} \qquad x_{\sigma}\equiv \frac{\hat\sigma}{\langle\sigma\rangle}\,.
\label{potsigma}
\ee
The requirement of having a trustable effective field theory sets
two new constraints on the value of the displacement $\hat\sigma$ from the minimum which take the form:
\bea
 &1)& \tau_4=\langle\tilde\tau_4\rangle +x\,\langle\tau_5\rangle \left(1+x_{\sigma}\right)^{4/3}>1
 \qquad \Leftrightarrow \qquad \left(1+x_{\sigma}\right)^{4/3}>
 \frac{1-\langle\tilde\tau_4\rangle}{x\,\langle\tau_5\rangle}\,, \label{constraint1} \\
 &2)& \tau_5=\langle\tau_5\rangle\left(1+x_{\sigma}\right)^{4/3}>1
 \qquad \Leftrightarrow \qquad \left(1+x_{\sigma}\right)^{4/3}>\frac{1}{\langle\tau_5\rangle}\,.
 \label{constraint2}
\eea
Therefore, from eq. (\ref{V}), we obtain the following potential for the modulating field, valid in regions of
$x_\sigma$  where the previous constraints hold:\footnote{We are neglecting possible up-lifting contributions
to the modulating field potential since they are irrelevant for the present discussion.}
\be
 U_{\rm mod} (x_\sigma)\simeq - A_0 p \ln\vo \left[\left(1 + \frac{1}{p \ln\vo}\right) \left(1 + x_{\sigma}\right)^{4/3}-1\right]\frac{W_0^2 }{\vo^{3+p (1 + x_{\sigma})^{4/3}}}\,.
\label{potfsig}
\ee
We point out that the potentials for the inflaton $\phi$ and the modulating field $\sigma$ are decoupled in this set-up,
and so the two fields can evolve independently in field space. This is also clear from the fact that $\langle\tau_5\rangle$
does not depend on $\langle\tau_1\rangle$, as we showed in section \ref{Fstab}.
The potentials are plotted in Fig \ref{fig1} for a representative choice of parameters.

In the following, we focus on the positive range for $x_\sigma$, where $U_{\rm mod}$ is bounded from above.
The quantities $A_0$ and $p$ depend on the microscopic properties of the model,
and for the parameter fit of eq. (\ref{secfit}) they take the values $A_0 \simeq 0.05$ and $p=1.5$.

We finally point out that the mass of the modulating field $\sigma$ for an arbitrarily large value of $x_\sigma$ is given by:
\be
 m^2_\sigma =\frac{1}{\langle\sigma\rangle^2}\frac{\partial^2 U_{\rm mod}}{\partial x_{\sigma}^2}\simeq
 \frac{W_0^2}{\vo^{2+p\left(1 + x_{\sigma}\right)^{4/3}}}\,.
\ee
As we have seen, $H^2\sim W_0^2 \vo^{-10/3}$. Thus if $p>\frac 43 \left(1+x_\sigma\right)^{-4/3}$
this mass is smaller than the Hubble parameter during inflation, and the field $\sigma$ is light during this phase.
Our choice of parameters ensures that this condition holds since we shall find that the requirement of generating enough
density perturbations fixes $x_\sigma\ll 1$. Hence we get $p > 4/3= 1.\bar{3}$ that is satisfied for our value $p= 1.5$.

The fact that $m_{\sigma}\ll H$ is an essential property to implement modulated reheating since it ensures that the field $\sigma$ is practically frozen at the classical level but it develops a scale invariant spectrum of
quantum fluctuations with amplitude set by the Hubble scale: $\delta\sigma \simeq H/2\pi$. The other crucial feature to realise the modulation mechanism is to have a $\sigma$-dependent inflaton coupling to visible degrees of freedom
so that the oscillations of $\sigma$ can indeed modulate the decay rate of the inflaton that gives rise to reheating.
We investigate this key property of our model in the next sections.

\subsection{Inflationary couplings to observable sectors}
\label{couplings}

An essential feature of the LVS framework is that we can
compute the volume scaling of the couplings between the moduli (among which the inflaton) and all the other visible or hidden degrees of freedom localised on D7-branes wrapped on internal four-cycles \cite{astro,Reheating,LVSatFiniteT}. This is a necessary ingredient for being able to calculate the inflaton decay rates into visible degrees of freedom. This allows us to understand the dynamics of reheating at the end of inflation \cite{Reheating}, in particular to determine whether a modulated reheating scenario can be successfully implemented in our set-up.

We focus on the decay channels of the moduli fields into light gauge bosons living on the D7-worldvolumes.
As described in \cite{astro,Reheating,LVSatFiniteT}, the corresponding couplings can be derived
from the moduli dependence of the tree-level gauge kinetic function given by eq. (\ref{gaugecoupling}).
The relevant kinetic terms read (writing ${\rm Re}(S)\equiv s$):
\be
 \mc{L}_{gauge}=- \left(\tau_i - h s\right) F_{\mu\nu}^{(i)}F^{\mu\nu}_{(i)}.
\label{Lgauge}
\ee
While the inflaton oscillates around its minimum after inflation ends, we consider the moduli as quantum fields $\delta \tau_i$ expanded around a classical value: $$\tau_i = \left( \langle\tau_i \rangle +\hat\tau_i\right)+\delta\tau_i\,,$$
where we take into account  the  possibility that classical VEVs are shifted from the minimum during cosmological evolution. Defining the canonically normalised field strength $G_{\mu\nu}^{(i)}$ as
\be
 G_{\mu\nu}^{(i)}=2\,F_{\mu\nu}^{(i)}\sqrt{\langle\tau_i\rangle+\hat\tau_i - h \langle s\rangle},
\label{redef}
\ee
we obtain (neglecting the piece involving $\delta s$ and taking $h\langle s \rangle \ll \tau_i$):
\begin{eqnarray}
 \mc{L}_{gauge}&=&-\frac{1}{4}\,G_{\mu\nu}^{(i)}G^{\mu\nu}_{(i)}-\frac{\delta\tau_i}
 {4 \left(\langle\tau_i\rangle+\hat\tau_i - h \langle s\rangle\right)}\,G_{\mu\nu}^{(i)}G^{\mu\nu}_{(i)} \notag \\
 &\simeq&-\frac{1}{4}\,G_{\mu\nu}^{(i)}G^{\mu\nu}_{(i)}-\left(1 +\frac{ h \langle s\rangle}{\langle\tau_i\rangle+\hat\tau_i}
 \right)\frac{\delta\tau_i}
 {4 \left(\langle\tau_i\rangle+\hat\tau_i\right)}\,G_{\mu\nu}^{(i)}G^{\mu\nu}_{(i)}.
\label{FmunuFmunu}
\end{eqnarray}
The moduli couplings to gauge bosons can be obtained
by expressing $\tau_i$ as a function  of the canonically normalised fields using the general field redefinitions
 of eqs. (\ref{TAU1})-(\ref{tau5canonica}), and expanding each field around its VEV as:
\be
 \phi= \langle\phi\rangle+\delta \phi,\qquad
 \chi_i=\langle\chi_i\rangle+\delta \chi_i \,\,\,\,i=2, 3, 4, \qquad
 \sigma= \langle\sigma\rangle+\hat \sigma\,+\delta \sigma\,.
\ee
In the previous expression, we included a possible shift from the minimum of the field $\sigma$
that is light both during inflation and the reheating process.
We instead consider the classical value of the inflaton $\phi$ as coinciding with its minimum in $\langle \phi \rangle$.
Indeed, as we know from the study of preheating in LVS \cite{Reheating}, the classical homogeneous condensate quickly
relaxes to its minimum, and inflaton self-quanta
get produced non-perturbatively. Then we can consider $\phi$ as
fixed at its minimum $\langle \phi \rangle$ during reheating.

As we discussed in section \ref{sec-consideration},
the visible sector fields live on a D7-brane wrapped around the divisor $D_{\rm vis}= v_4 D_4+v_5 D_5$
whose volume is given by $\tau_{\rm vis}=v_4\tau_4+v_5\tau_5$.
Therefore, the couplings of the moduli fields to the visible sector
gauge bosons living on $D_{\rm vis}$ can be obtained from \cite{BCGQTZ, astro}:
\be
 \frac{\delta \tau_{\rm vis}}{4\,\tau_{\rm vis}}=
 u_1\,\delta\phi
 + \frac{1}{2\sqrt{2}}\,u_2\,\delta\chi_2
 -\frac 38\langle\chi_3\rangle \,\delta\chi_3
  +\frac 13 \,u_4 \,\delta\chi_4
 +\frac 13 \,u_5 \,\delta\sigma,
\ee
where the parameters $u_i$ can be obtained from eqs. (\ref{tau4canonica}) and (\ref{tau5canonica}).
Given that their expression is quite long, we relegate it to appendix \ref{app-parameters}.
This information is what we need to express
the following couplings (denoting as $\gamma$ the visible sector gauge bosons and introducing factors of $M_P$ to recast the correct dimensions):
\footnote{The coupling of $\delta\chi_3$ to visible sector gauge bosons is induced by
the mixing between $\vo$ and $\tau_3$ induced by subleading string loop effects, and it scales as
$\lambda_{\delta\chi_3\gamma \gamma}\sim \langle\chi_3\rangle/M_P^2 \sim 1/\left(\sqrt{\vo} M_P\right)$.}
\be
 \lambda_{\delta\phi\gamma \gamma}=\frac{u_0 u_1}{M_P},\quad
 \lambda_{\delta\chi_2\gamma \gamma}=\frac{1}{2\sqrt{2}}\frac{u_0 u_2}{M_P}, \quad
 \lambda_{\delta\chi_4\gamma \gamma}=\frac 13 \frac{u_0 u_4}{M_P},\quad
 \lambda_{\delta\sigma\gamma \gamma}=\frac 13 \frac{u_0 u_5}{M_P}.
\label{Modcoupl}
\ee
The most important one is the coupling of the inflaton field $\phi$ to visible sector gauge bosons since it is this coupling which is responsible for the perturbative decay of the inflaton that reheats the visible sector degrees of freedom. The interesting observation is that this coupling depends on the
value of the field $\sigma$, and so it can be modulated by the frozen quantum oscillations of this light field. In fact, it scales schematically as:
\be
 \lambda_{\delta\phi\gamma \gamma }(x_\sigma) =
 \left[1+\frac{c_1}{1+ c_2 \left(1+x_\sigma\right)^{4/3}}\right]
 \frac{c_0}{\langle\vo\rangle^{1/3} M_P},
 \label{l144n}
\ee
where the $c_i$'s are positive $\mc{O}(1)$ constants,
that can be computed using the formulae of section \ref{sect-can}. They turn out to be given by:
\be
 c_0 = u_1\,\vo^{1/3},\qquad
 c_1 = \frac{h\left(\mc{F}_{\rm vis}\right)}{g_s v_4\langle\tilde\tau_4\rangle}
 =\frac{3 {\mathbf b}_4 (2 f_4 + v_4)^2}{ 4 g_s \langle\tilde\tau_4\rangle}>0,
 \qquad c_2=\frac{\left(x\,v_4+v_5\right)\langle\tau_5\rangle}{v_4\langle\tilde\tau_4\rangle}\,.
\label{valc1c2}
\ee
We stress that only the sign of $c_1$ is fixed and the two constants $c_1$ and $c_2$
depend on the underlying parameters of our model. Choosing:
\be
k_{444}=101\qquad \text{and}\qquad k_{445}=k_{455}=k_{555}=100\,,
\ee
which give:
\be
{\mathbf a}_4={\mathbf a}_5 = \left(\frac{50}{3}\right)^{1/3}\quad \Leftrightarrow \quad x=1
\qquad \text{and} \qquad {\mathbf b}_4=\frac 16\,,
\ee
we obtain:
\be
c_1 \simeq 9.8\cdot 10^{-4}\qquad\text{and}\qquad c_2\simeq -1.004\,.
\ee

The decay rate of the inflaton field $\phi$ into MSSM gauge bosons can be written as:
\be\label{exprGam}
 \Gamma_{\delta \phi\to\gamma \gamma }\,=\,\frac{N_g\,
 \lambda^2_{\delta\phi\gamma \gamma }\,m_\phi^3}{64 \pi},
 \ee
where the coupling $\lambda_{\delta\phi\gamma \gamma }$ is given by eq. (\ref{l144n}),
$m_\phi$ is the mass of the inflaton around its minimum, and $N_g=12$ is the total number of MSSM gauge bosons.

We conclude pointing  out that our brane set-up is such that all the four-cycles
are wrapped by stacks of D7-branes. Hence we have also to take into account the moduli couplings
to gauge bosons living on all these hidden sectors. Notice that in this case we are interested
at the volume scaling of these couplings but not at their exact field dependence since they are not controlling reheating.
The volume scaling of the moduli couplings to all gauge bosons present in the model is listed in Table 1.
The only issues that remain to be checked are  that the inflaton energy is not dumped into hidden,
instead of visible, degrees of freedom at the end of inflation \cite{Reheating}, and that the
 modulating field decays before the inflaton. We discuss these points in the next section.

\begin{table}[ht]
\begin{center}
\begin{tabular}{c||c|c|c|c|c}
  & $\delta\phi$ & $\delta\chi_2$
  & $\delta\chi_3$ & $\delta\chi_4$ & $\delta\sigma$ \\
  \hline\hline
  \\ & & & & \vspace{-0.9cm}\\
  $(F_{\mu \nu}^{(1,2)} F^{\mu \nu}_{(1,2)})$
  & $\frac{1}{M_P}$
  & $\frac{1}{M_P}$
  & $\frac{1}{\vo^{1/2}M_P}$
  & $\frac{1}{\vo^{1/2}M_P}$
  & $\frac{1}{\vo^{1/2}M_P}$ \\
  \hline
  \\ & & & & \vspace{-0.9cm}\\
  $(F_{\mu \nu}^{(3)} F^{\mu \nu}_{(3)})$
  & $\frac{1}{\vo^{1/3} M_P}$
  & $\frac{1}{M_P}$
  & $\frac{\vo^{1/2}}{M_P}$
  & $\frac{1}{\vo^{1/2}M_P}$
  & $\frac{1}{\vo^{1/2}M_P}$ \\
  \hline
  \\ & & & & \vspace{-0.9cm}\\
  $(F_{\mu \nu}^{\rm vis} F^{\mu \nu}_{\rm vis})$
  & $\frac{1}{\vo^{1/3} M_P}$
  & $\frac{1}{M_P}$
  & $\frac{1}{\vo^{1/2}M_P}$
  & $\frac{\vo^{1/2}}{M_P}$
  & $\frac{\vo^{1/2}}{M_P}$
\end{tabular}
\end{center}
\caption{Volume scaling of the moduli couplings to all gauge bosons in the model.
We denote as $F_{\mu \nu}^{(i)}$ the field strength associated with the gauge bosons living on the cycle $D_i$, $i=1,2,3$.}
\end{table}

\subsection{Reheating}
\label{sec-reheating}

In this section we shall study how reheating takes place,
ensuring that the inflaton does not dump all its energy to
hidden, instead of visible, degrees of freedom at the end of inflation \cite{Reheating}.
Moreover we shall also state the conditions under which the modulating field decays before the inflaton
without giving rise to isocurvature contributions.

At the end of inflation, the inflaton field $\phi$ coherently oscillates around its minimum behaving as a classical homogeneous condensate.
In the case of \emph{blow-up inflation} \cite{kahlerinfl}, the inflaton energy gets transferred to the visible sector via
a first period of preheating \cite{Preheating} characterised by a violent non-perturbative production of inflaton quanta whose perturbative decay subsequently gives rise to reheating \cite{Reheating}. In our case, given that the shape of the inflationary potential for \emph{fibre inflation} \cite{fiberinfl}
is much less steep than the one for \emph{blow-up inflation}, we expect a different behaviour for preheating without such a violent non-perturbative production of inflaton inhomogeneities. It might even be that in this case all the dynamics is purely perturbative.

Due to the presence of many hidden sectors in our model,
an important issue to check in order to understand reheating,
is whether the inflaton $\phi$ decays first to visible or hidden degrees of freedom. As can be seen from Table 1, $\phi$ couples to the hidden gauge bosons living on $D_1$ and $D_2$ more strongly than to the visible gauge bosons living on $D_{\rm vis}$.
Thus the inflaton would naively dump all its energy to hidden, instead of visible, degrees of freedom.

This problem can be solved if the inflaton decay to hidden
degrees of freedom is kinematically forbidden. In \cite{Reheating},
it has been shown that this is indeed the case if the hidden sector consists
of a pure $\mc{N}=1$ SYM theory that develops a mass gap.
From now on, we shall take this as a viable solution.
Hence we shall assume that proper non-trivial fluxes can fix the D7-deformation moduli
associated with the non-rigidity of $D_1$ and $D_2$,
that would prevent the generation of non-perturbative effects.
 We also point out that moduli stabilisation is not affected by these new
non-perturbative corrections in $\tau_1$ and $\tau_2$ since they are subdominant with respect to the string loops in the regime of interest where both of these four-cycles are large. We have to consider a pure SYM theory also on $D_3$ otherwise too much hidden sector dark matter would be produced by the simultaneous inflaton decay to visible and hidden degrees of freedom living on $D_3$ \cite{Reheating}.

Under all these constraints on the hidden sector model building, reheating can take place with a temperature of the order:
\be
 \Gamma_{\phi\to \gamma\gamma}\simeq \lambda_{\phi\gamma\gamma}^2m_{\phi}^3
 \simeq \frac{M_P}{\vo^{17/3}}
 \,\,\,\,\Rightarrow\,\,\,\, T_{RH}\simeq\sqrt{\Gamma_{\phi\to\gamma\gamma} M_P}\simeq \frac{M_P}{\vo^{17/6}}.
\label{TRH4}
\ee
The requirement of having a reheating temperature above Big Bang Nucleosynthesis, {\em i.e.}~$T_{RH}> 1$ MeV, sets an upper bound on the value of the overall volume $\vo$ of the order $\vo\leq 10^7$ in string units. It is crucial to notice that this bound still allows us to choose values of the volume, $\vo\geq 10^3$, for which the standard mechanism for the generation of the density perturbations can be made negligible.
From now on, we shall therefore work in the window $10^3 \leq\vo\leq 10^7$, consistent with the parameter choices we made in the previous sections.

Lastly we point out that for an appropriate choice of the parameter $p$, the modulating field $\sigma$ decays before the inflaton ensuring that there are no curvaton-like contributions to the generation of the density perturbations from the decay of $\sigma$. In fact, comparing the two decay rates we obtain:
\be
 \Gamma_{\sigma\rightarrow \gamma \gamma}
 \simeq\lambda_{\sigma\gamma\gamma}^2 m_{\sigma}^{3}\simeq \frac{M_P}{\vo^{2+\frac{3p}{2}}},\,\,\,\,
 \Rightarrow\,\,\,\,
 \Gamma_{\sigma\to\gamma\gamma}>\Gamma_{\phi\to \gamma\gamma}\,\,\,\,
 \Leftrightarrow\,\,\,\, p<\frac{22}{9}.
\label{Gsigma}
\ee
Notice that this upper bound on $p$ still allows us to choose values of this parameter, $p>4/3$, for which the field $\sigma$ is lighter than $H$ during inflation and so can act as a modulating field. From now on, we shall therefore work in the window of values for $p$, $4/3<p<22/9$, or equivalently $1.\bar 3 < p < 2. \bar 4$.

We finally point out that we derived the decay rates (\ref{TRH4}) and (\ref{Gsigma}) assuming that the modulating field
is close to its minimum, {\em i.e.}~$x_\sigma\ll 1$. If this is not the case, a possible worry is that
there might be a second period of inflation driven by $\sigma$.
However given that $\sigma$ decays before $\phi$ this is not the case.
In fact, during the initial period of inflation driven by $\phi$, if $x_\sigma \gtrsim 1$,
$\sigma$ is classically frozen in a region of field space far from its minimum.
Then when inflation ends, $\sigma$ starts to slowly roll down its potential
but is does not drive inflation because all the energy is stored in the inflaton self-quanta
which have been produced non-perturbatively at preheating.
If such quanta decayed before $\sigma$, then $\sigma$ would have enough time to dominate the energy density and drive
a second period of inflation but, as we have shown in (\ref{Gsigma}), this does not happen if $p<22/9$.
Hence for such a value of $p$, while $\sigma$ is slow-rolling, the $\phi$ quanta have not decayed yet.
Then $\sigma$ ends slow-rolling, decays and later on the $\phi$ quanta decay.
Thus we need to make sure that $\sigma$ decays after $\phi$ not just to neglect
curvaton-like contributions but also to ensure the absence of second period of inflation driven by $\sigma$
in the case when $x_\sigma \gtrsim 1$.

\section{Curvature perturbations from modulated reheating}
\label{sec-modreh}

This section describes the generation of primordial fluctuations,
first in a general context and then in the specific string-inflation scenario considered above.

\subsection{Modulated reheating in LVS inflationary scenarios}
\label{sect-gendis}

Before presenting the details of how curvature perturbations are produced in our set-up,
we open this section with a more general discussion on modulated reheating in LVS.

The best moduli candidates to play the r\^ole of inflaton and modulating fields can be identified by analysing
the conditions that these fields have to satisfy to yield a natural realisation of the modulation mechanism:
\begin{enumerate}
\item The modulating field must be lighter than the Hubble scale $H$ during inflation.

\item The contribution of the modulation mechanism to the spectrum  of  density perturbations must be larger,
or at most equal to the one of the standard mechanism: ${\cal P}_{\zeta_{\rm mod}} \gtrsim  {\cal P}_{\zeta_{\rm stand}}$.
These two contributions are given by (at leading order and in the approximation of instantaneous reheating, {\em i.e.}~$\Gamma\gg H$):
\be
{\cal P}^{1/2}_{\zeta_{\rm stand}}
=\frac{1}{2\sqrt{2}\pi} \frac{H}{M_P\sqrt{\epsilon}}\,,\qquad
{\cal P}^{1/2}_{\zeta_{\rm mod}}
=\frac{1}{6} \frac{\delta\Gamma}{\Gamma}\,,
\ee
where $\Gamma$ is the decay rate modulated by the fluctuations of the modulating field.
\end{enumerate}

All the decay rates in our model are functions of the mass of the decaying particle $m$ and its coupling $\lambda$ to gauge bosons, which, in turn, can be functions of the modulating field. As we showed in the previous sections, both the fibre modulus $\phi$ and the blow-up mode $\sigma$ can be made lighter than $H$ during inflation, so both of these fields satisfy the first condition.

However the field $\phi$ does not satisfy the second one due to the fact that its canonical normalisation makes both its mass (see \ref{Inflpot}) and its coupling to gauge bosons (see \ref{Modcoupl} and appendix \ref{app-parameters}) depend exponentially on $\phi$.
This implies that ${\cal P}^{1/2}_{\zeta_{\rm mod}} \simeq \delta \phi\simeq  H/M_P$, and so the standard mechanism can never be beaten for the typical inflationary requirement $\epsilon\ll 1$. \footnote{Inflaton couplings to photons which have an exponential dependence on the modulating field have been
discussed in \cite{Japan} where the authors pointed out that in this case the decay of the inflaton to the modulating field plus two photons dominates over the decay to just two photons. Consequently, the modulating field can behave as dark radiation. It is interesting to notice that our string embedding does indeed include inflaton couplings which depend exponentially on the modulating field as assumed in \cite{Japan}. However in order to beat the standard mechanism, one needs a dependence of the form $e^{\phi/M}$ where $M \ll M_P$ whereas in our case we find $M \simeq M_P$.}

On the other hand, the different form of the canonical normalisation of a fibre divisor and a blow-up mode, gives rise to a non-exponential dependence of the inflaton coupling to gauge bosons (\ref{l144n}) on the blow-up mode $\sigma$, rendering this field a more promising modulating field candidate.

The mass of the inflaton field (see \ref{Inflpot}) does not depend on $\sigma$, and so we need only to take into account the dependence of the coupling (\ref{l144n}) on $\sigma$, obtaining:
\be
 \zeta_{\rm mod} = -\frac{1}{6\pi}\frac{H}{\lambda}\frac{\partial\lambda}{\partial\hat\sigma}
 =\frac{2 c_1 c_2 (1 + x_\sigma)^{1/3}}{9\pi
  \left[1 + c_2 \left(1 + x_\sigma\right)^{4/3}\right]
 \left[1 + c_1 + c_2 \left(1 + x_\sigma\right)^{4/3}\right]} \frac{H}{\langle\sigma\rangle M_P}\,.
\label{zm}
\ee
From the canonical normalisation (\ref{smalltaus}) we realise that $\langle\sigma\rangle \simeq \langle\tau_5\rangle^{3/4} \vo^{-1/2}$,
and so for $x_\sigma\sim c_1\sim c_2\sim \mc{O}(1)$, (\ref{zm}) reduces to:
\be
 \zeta_{\rm mod}\sim \sqrt{\vo}\frac{H}{M_P}\sim \frac{H}{M_s}\,,
\label{zbig}
\ee
regardless of the strength of the modulated coupling!
This is due to the fact that $M_s$ comes from the value of the VEV of the canonically normalised blow-up mode $\sigma$ even if the modulating coupling (\ref{l144n}) is weak since it scales as $\lambda \sim 1/(\vo^{1/3} M_P)$. This result implies that the ratio between the amplitude of the density perturbations generated by the two mechanisms behaves as:
\be
 R\equiv \frac{\zeta_{\rm mod}}{ \zeta_{\rm stand}} \sim \sqrt{\vo\,\epsilon},
\ee
and so in order to have $R\geq 1$ we need also $\epsilon$ relatively large.

The value of $\epsilon$ depends on the nature of the inflaton field. This is the reason why we chose the inflaton as a fibre divisor. In fact, $\epsilon\sim 10^{-4}$ in \emph{fibre inflation}
\cite{fiberinfl} whereas $\epsilon\sim 10^{-12}$ in \emph{blow-up inflation} \cite{kahlerinfl} where the inflaton is another blow-up mode.
With respect to this, let us present a brief proof of the fact that $\epsilon$ in \emph{blow-up inflation} is so small that the modulation mechanism can never beat the standard one. Even assuming a $\zeta_{\rm mod}$ that is the largest possible, {\em i.e.}~$R \sim \sqrt{\vo\,\epsilon}$,
since $\epsilon$ goes like \cite{kahlerinfl}:
\be
 \epsilon \sim \vo^3 e^{-2 a \tau_{\rm inf}}\text{ \ \ with \ \ }e^{- a \tau_{\rm inf}}\ll \vo^{-2}\text{ \ \ to have \ \ }\eta\ll 1,
\ee
it turns out that $\epsilon^{1/2}\ll\vo^{-1/2}$ which implies $R\ll 1$.

We showed why we chose $\phi$ as our inflaton field and $\sigma$ as our modulating field. The fact that in \emph{fibre inflation} $\epsilon\sim 10^{-4}$ \cite{fiberinfl}, implies that in order to get $R\gtrsim 1$ we have to focus on volumes of the order $\vo \gtrsim 10^3$ in string units. This is indeed the range that we are considering in our set-up. We point out that the only problem to use a blow-up mode as a modulating field, is to render this field lighter than $H$ during inflation given that, 
in the standard LVS, all the blow-up modes nearby their minima are naturally heavier than the volume mode that sets the scale of the scalar potential.
However we showed that this is indeed possible using poly-instanton corrections to the superpotential.

We finally stress that our model represents the first
realisation from string theory of a standard modulated reheating scenario where both inflation and the domination over the standard mechanism for the generation of the density perturbations are achieved without requiring any fine-tuning of the underlying parameters.

\subsection{Non-gaussianity}\label{sect:nongau}

The fact that the decay rate of the inflaton field $\phi$ to visible degrees of freedom depends on a light field $\sigma$, has important implications for the generation of adiabatic curvature perturbations $\zeta$.  Since $\sigma$ is light during inflation, it acquires quantum fluctuations that render its value at horizon exit slightly different in different parts of the Universe. This implies that reheating can occur at different times in different places, depending on the value of $\sigma$. A consequence of this phenomenon is that the isocurvature fluctuations associated with the field $\sigma$ can be converted into adiabatic
curvature fluctuations $\zeta$. This is the basic idea of the {\it modulated  reheating} scenario \cite{DGZ1,Kofman,DGZ2} to produce the curvature fluctuations in a way that is alternative to the standard mechanism. This way to produce the curvature fluctuations is particularly interesting, since it can lead to non-gaussianity of local form, with an amplitude much larger with respect to the non-gaussianity normally generated in single field inflation,  that are controlled by slow-roll parameters.

In this section, we study in detail the properties of the curvature fluctuations in our set-up. Let us focus on the case in which  the modulation of the inflaton decay rate by the field $\sigma$ is entirely responsible for producing the adiabatic curvature fluctuation $\zeta$, whose power spectrum matches the COBE normalisation. This is the situation we are considering here, since in our set-up (see eq.~(\ref{inflprop})) the inflaton contribution to the spectrum of curvature perturbations is negligible.

In our set-up the curvature perturbations are produced by the fluctuations of the decay rate of the inflaton into visible sector fields. Besides the original works on modulated reheating, a careful analysis of how adiabatic perturbations and  non-gaussianity  are generated  is developed for example in \cite{ZaldarriagaNG, Suyama:2007bg, Ichikawa:2008ne}. At the linear level, the curvature perturbation is proportional to the spatial variations of the inflaton decay  rate:
\be
 \zeta \,=\,-\frac16 \,\frac{\delta \Gamma}{\Gamma},
\ee
where the overall factor $-1/6$ is calculated by taking into account the dynamics of the  inflaton  field
between the end of inflation and reheating. This value for the overall factor is found in the limit in which the ratio between the inflaton decay rate and the Hubble parameter is very large, a situation that is satisfied in our context.

Going beyond the linear level, one takes a higher order expansion for the curvature perturbations as follows \cite{Suyama:2007bg}:
\be\label{expzeta}
 \zeta \,=\,-\frac16 \,\frac{\delta \Gamma}{\Gamma}+\frac{1}{2}\,\left(\frac16
 \,\frac{\delta \Gamma}{\Gamma}\right)^2-\frac{1}{3!}\,\left(\frac16
 \,\frac{\delta \Gamma}{\Gamma}\right)^3+\cdots\,.
\ee
The spatial fluctuations in the inflaton decay rate are associated to the spatial fluctuations $\delta \sigma$ of the light field $\sigma$ during inflation. It is  convenient to make an expansion of this quantity in power series of $\delta \sigma$
as follows:
\be\label{expdgam}
 \delta \Gamma\,=\,\,\Gamma_\sigma \delta \sigma+\frac{1}{2} \Gamma_{\sigma \sigma}
 \delta \sigma^2+\frac{1}{3!} \Gamma_{\sigma \sigma \sigma}\delta \sigma^3\,+\cdots\,.
\ee
We regard $\delta \sigma$ as corresponding  to the random fluctuation of a light scalar during an inflationary quasi-de Sitter phase, and so we assume that $\delta \sigma$ is a Gaussian random variable. The amplitude of the associated two point function is controlled by the Hubble parameter $\langle \delta \sigma^2 \rangle \,\propto\,H^2/(2 \pi)^2$; odd-point
functions and the remaining connected even point functions vanish. Plugging the expression for (\ref{expdgam}) into eq.~(\ref{expzeta}), we find an expression for $\zeta$ as an expansion in $\delta \sigma$ that assumes the typical form leading to {\it local non-gaussianity} developing  after horizon exit \cite{komatsu-spergel}. With this form for the curvature perturbation $\zeta$, it is straightforward to calculate the two, three and four point functions of the curvature perturbations to the power spectrum, bispectrum and trispectrum:
\bea
 \langle \zeta_{k_1} \zeta_{k_2}\rangle&=&\left( 2 \pi \right)^3\,\delta(\vec k_1 +
 \vec k_2)\,{\cal P}_{\zeta} (k_1)/k_1^3, \\
 \langle \zeta_{k_1} \zeta_{k_2} \zeta_{k_3}\rangle&=&\left( 2 \pi \right)^3\,\delta(\vec k_1 +
 \vec k_2+
 \vec k_3)\,B_{\zeta} (k_1, k_2, k_3), \\
 \langle \zeta_{k_1} \zeta_{k_2} \zeta_{k_3} \zeta_{k_4}
 \rangle&=&\left( 2 \pi \right)^3\,\delta(\vec k_1 +
 \vec k_2+\vec k_3+
 \vec k_4)\,T_{\zeta} (k_1, k_2, k_3,k_4).
\eea
We finally express the bispectrum and trispectrum in terms of $f_{\rm \NL}$, and of $g_{\rm \NL}$ and $\tau_{\rm \NL}$. One finds that the power spectrum depends on  the first derivative of $\Gamma$ with respect to $\sigma$:
\be\label{powermodul}
 {\cal P}_\zeta=
 \left( \frac{H }{12 \pi}\frac{\Gamma_\sigma}{\Gamma}\right)^2 \,,
\ee
while the non-gaussianity parameters depend also on higher derivatives of $\Gamma$ \cite{Suyama:2007bg}:\footnote{In order to find this result, we have taken the case in which $\Gamma/H \ll 1 $, as shown in \cite{Suyama:2007bg}.}
\bea
 f_{\rm \NL}&=&
 5\left( 1-\frac{\Gamma_{\sigma  \sigma} \Gamma}{\Gamma_\sigma^2}\right),
 \\
 g_{\rm \NL}&=&\frac{50}{3}\left(2-3 \frac{  \Gamma_{\sigma
 \sigma} \Gamma}{\Gamma_\sigma^2}+\frac{
 \Gamma_{\sigma \sigma
 \sigma} \Gamma^2
 }{\Gamma_\sigma^3} \right), \\
 \tau_{\rm \NL}&=&\left( \frac65 f_{\rm \NL}\right)^2.
\eea
We can then appreciate why  non-gaussianity has the opportunity to  be much larger than the slow-roll parameters  in this context: they depend on quantities (derivatives of $\Gamma$) that  do not directly control the inflationary dynamics, nor other aspects of the Universe homogeneous evolution. Consequently, they do not have to satisfy the stringent upper bounds associated with the dynamics of inflation, as happens for example to the non-gaussianity generated by the inflaton field in standard single-field models of inflation, that are controlled by slow-roll parameters.

Plugging the expression for the inflaton decay rate obtained in eq. (\ref{exprGam}), and using the $\sigma$-dependence of the coupling $\lambda_{\delta \phi \gamma \gamma}$ of eq. (\ref{l144n}), we get the following expressions  for ${\cal P}_\zeta$, $f_{\rm \NL}$ and $g_{\rm \NL}$ in terms of the underlying parameters:
\bea
 {\cal P}_\zeta &=& \frac{64\, c_1^2\, c_2^2\, \sqrt{y_{\sigma }}}{9 (1 + c_2 y_{\sigma})^2 (1 + c_1 + c_2 y_{\sigma})^2} \left(\frac{H}{12\pi\langle\sigma\rangle}\right)^2\,, \label{Pz} \\
 f_{\NL}&=&\frac{5 \left[1 + c_1 - 3 c_2 y_{\sigma} \left(2  +  c_1\right)-
 7 c_2^2 y_{\sigma}^2\right]}{8 \,c_1\, c_2\, y_{\sigma }}\,, \label{fnl}\\
 g_{\NL}&=&-\frac{25}{48\, c_1^2\, c_2^2\, y_{\sigma }^2} \left[1 + 2 c_1 + c_1^2 + 8 (2 + 3 c_1 + c_1^2) c_2 y_\sigma \right. \nn \\
 && \left.- 3 (2 + 2 c_1 + 3 c_1^2) c_2^2 y_\sigma^2 - 28 (2 + c_1) c_2^3 y_\sigma^3 -
 35 c_2^4 y_\sigma^4\right]. \label{gnl}
\eea
where $y_\sigma \equiv  \left(1+x_\sigma\right)^{4/3}\,=\,  (1+\hat\sigma /\langle\sigma\rangle)^{4/3} $,
and $\hat \sigma$ is the classical shift of the field $\sigma$ from its minimum (see discussion after eq. (\ref{seconda})).
The shift $\hat \sigma$ can be different than zero: since this field is light during inflation,
its classical value is frozen during this phase to a position that does not necessarily correspond to its minimum.

We have also to satisfy the constraints in equations (\ref{constraint1}) and (\ref{constraint2}).
These requirements set stringent constraints on the parameters of the model,
and provide very definite predictions on the amount of non-gaussianity produced in this cosmological scenario.
On the other hand,
the fact that the non-gaussianity parameters and the quantity $U_{\rm mod}$ are
non-trivial functions of $\hat \sigma$ enriches
the phenomenological features of our model. For example, non-gaussianity can exhibit sizeable scale dependence
\cite{Byrnes:2010ft}, that might be used for further characterising the predictions of our set-up.

In order to generate a power spectrum of curvature fluctuations that matches the COBE normalisation,
given our choice of parameters in  equation  \ref{secfit},
 we need to choose:
\be
x_\sigma \simeq 9.5\cdot 10^{-4}\,,
\ee
which gives rise to the following predictions for the   parameters controlling local non-gaussianity:
\be
 f_{\rm \NL}\simeq 23.84\,, \qquad
 g_{\rm \NL}\simeq 570.5\,, \qquad
 \tau_{\rm \NL} \simeq 818.5 \,.
\ee

The value of $f_{\rm \NL}$ lies very well within the allowed window $-10<f^{local}_{\rm \NL}<74$ at $95\%$ confidence level from WMAP7 data \cite{WMAP7}. 
Moreover, our prediction for $f_{\rm \NL}$ is large enough to be testable by the Planck satellite which will
be able to measure this parameter with an accuracy of the order $\Delta f_{\rm \NL}\simeq 5$ \cite{KomatsuReview}.
On the other hand, $g_{\rm \NL}$ and $\tau_{\rm \NL}$ are too small to be detected by the Planck
satellite (even if such values for $\tau_{\rm \NL}$
might be observable by future  CMB experiments as EPIC \cite{EPIC}).
Let us conclude by saying that these values of the non-gaussianity parameters are sensitive to our parameter choice, 
and so to the geometrical properties of our set-up. By changing the values of the quantities in formula (\ref{secfit}) one can obtain different predictions
for non-gaussianity: typically, the resulting value of $f_{\rm \NL}$ is of order a few.~\footnote{Still detectable
by future measurements of the large-scale structure of the Universe that will achieve an accuracy of the order $\Delta f_{\rm \NL}\simeq 1$
(see \cite{NGreview} for a review).} On the other hand, let us emphasise again that 
the explicit example above shows concretely that 
non-gaussianity, in our set-up, can be large enough to be observable. This opens up the possibility 
to have explicit string theory models of cosmology which are testable in the near future.

\section{Conclusions}

In summary, this paper provides a first realisation of the modulation mechanism within an explicit string inflationary construction, a process which proved surprisingly difficult to achieve given how well-matched the modulation mechanism is ({\em i.e.}~light fields, VEVs controlling couplings and masses, {\em etc.}) to the properties of string theory.

Together with the recent realisation of the curvaton mechanism \cite{BCGQTZ}, this provides two instances of string-inflationary models for which primordial fluctuations are generated by converting inflationary isocurvature fluctuations into adiabatic fluctuations during the post-inflationary reheating epoch. But given the apparent ubiquity of light scalar fields during string inflation, these are unlikely to be the last two to be found. More likely, the predominance of standard inflaton-generated fluctuations among string-inflation models is a `lamp-post effect,' reflecting the process through which these models were sought.

The class of inflationary model used here --- and in \cite{BCGQTZ} --- is driven by the time-evolution of an extra-dimensional modulus, within the context of the LARGE Volume Scenario (LVS) for stabilising string vacua \cite{LV}. This is a particularly useful scenario for these purposes, providing as it does a simple and robust mechanism for obtaining exceptionally light scalars \cite{GenAnalofLVS}, whose masses are suppressed relative to other moduli (and to the inflationary Hubble scale \cite{fiberinfl}) by inverse powers of the large volume. At present, the greatest weakness of these models is our relatively poor understanding of the details of their potential shape, due to the lack of explicit string loop calculations. But the main features, such as the ubiquity of light scalars they predict, depend only on the properties of the tree-level potential and so do not depend on these details.

As ever, although the resulting model is somewhat complicated, several things are gained by embedding a generic inflationary mechanism --- like the modulation and curvaton mechanisms for generating primordial fluctuations --- into an explicit UV completion, like an explicit string vacuum. Not least among these is the existence proof that the mechanisms can really be viable, given the extreme sensitivity inflation has towards the trickle-down of irrelevant (in the RG sense) gravitational effective interactions to the moderately low energies appropriate to inflation. This includes in particular the ease with which effective `Planck slop' interactions can compete with the inflaton self-couplings, which must be arranged to be small by construction in order to achieve an inflationary slow roll \cite{StrInfRevs}.

But other benefits of having a UV completion are also likely to accrue over the somewhat longer term. Among these is the ability to understand the interplay among the conflicting demands put on the microscopic couplings by the sometimes conflicting requirements of successful inflationary phenomenology. In particular, a UV completion is required in order to evaluate the competition among the various fluctuation-generation mechanisms that are at play, to see under which circumstances each can dominate. This is a special case of the broader observation that a proper understanding of reheating requires a complete theory of {\em all} of the degrees of freedom that are present at the relevant energies. (A fuller knowledge than just the couplings between the inflaton and observable low-energy fields is required because without knowing all of the degrees of freedom it is impossible to tell whether other channels might too successfully drain away any heat generated by a putative reheating source \cite{Reheating}; what has been called `The Parable of the Canadian Winter.')

Ultimately one hopes having a UV completion to general low-energy mechanisms is a small step on the way towards a longer-term goal: to find a string vacuum that does it all. That is, the challenge of having a multitude of string solutions in the landscape is not that we have too many ways to describe what we find around ourselves; rather, it is that we have not yet found even one way to describe everything we see. The challenge is to find the proverbial needle in the haystack. We believe our best hope to remedy this situation is to identify string `modules': that is, explicit string vacua whose low-energy limits exhibit a host of desirable phenomenological features: the Standard Model, successful inflationary cosmology and reheating, an understanding of dark matter and dark energy, and so on. All of these can put conflicting conditions on the same UV couplings \cite{InfvsSUSY}, that must simultaneously account for them all. We have seen that though it is possible to obtain large non-gaussianities from string inflation, it is not generic. If non-gaussianities are discovered they would put strong constraints on the proper string compactification eliminating many, but as we have shown, not all of them.

UV completions are also useful for mapping out what can be possible, and this is likely to be a more pressing priority should primordial non-gaussianity be discovered. Such a discovery would immediately precipitate a search for the kinds of mechanisms towards which the evidence points, and by extension also for the kinds of fundamental theories with which they are compatible. Until recently the discovery of observably large, local non-gaussianity could have been inferred as evidence against string theoretical inflationary models, and by extension  giving preference for its alternatives \cite{Cyclic}, given that these alternatives could generate such a non-gaussianity \cite{CyclicNG} while the only known string inflation-generated sources were not of the local type \cite{SFnongauss}. This paper, and ref.~\cite{BCGQTZ}, are first arguments against drawing such a conclusion, and are the first steps towards a broader survey of the string-theoretic options for generating primordial fluctuations.

\section*{Acknowledgments}

We thank Marta G\'omez-Reino for collaboration in the early stages of this work.
We would also like to thank Paolo Creminelli, Lev Kofman, Francisco Pedro, Roberto Valandro
 and David Wands for useful discussions,
and Tony Riotto for encouraging us to explore non-standard ways to generate primordial fluctuations.
Various combinations of us are grateful for the the support of, and the pleasant environs provided by, the Abdus Salam International Center for Theoretical Physics (ICTP), Perimeter Institute and Cook's Branch. CB's research was supported in part by funds from the Natural Sciences and Engineering Research Council (NSERC) of Canada. Research at the Perimeter Institute is supported in part by the Government of Canada through Industry Canada, and by the Province of Ontario through the Ministry of Research and Information (MRI). GT is supported by an UK STFC Advanced Fellowship ST/H005498/1.

\begin{appendix}

\section{Kinetic terms for the moduli}\label{app-kin}

In this section we investigate the field redefinitions needed to
put the kinetic terms into canonical form. The starting point in
the regime of interest is the K\"ahler metric for the moduli,
which is given by the following symmetric matrix:
\begin{equation}
K_{i\bar{\jmath}}^{0}=\frac{3}{8}\left(
\begin{array}{ccccc}
\frac{2}{3\tau_1^2} & \frac{2\alpha^2\left(\gamma_3\tau_3^{3/2}+\gamma_5\tau_5^{3/2}
+\gamma_4\tilde{\tau}_4^{3/2}\right)}{3\vo^2\sqrt{\tau_1}} & -\frac{\alpha\gamma_3\sqrt{\tau_3}}{\tau_1\vo}
& -\frac{\alpha \gamma_4\sqrt{\tilde{\tau}_4}}{\tau_1\vo} &
-\frac{\alpha\left(\gamma_5\sqrt{\tau_5}- \gamma_4 x \sqrt{\tilde{\tau}_4}\right)}{\tau_1\vo} \\
^{\prime \prime } & \frac{4\alpha^2\tau_1}{3\vo^2} & -\frac{2\alpha^2\gamma_3\sqrt{\tau_1\tau_3}}{\vo^2}
& -\frac{2\alpha^2 \gamma_4\sqrt{\tau_1\tilde{\tau}_4}}{\vo^2} &
-\frac{2\alpha^2\sqrt{\tau_1}\left(\gamma_5\sqrt{\tau_5}-\gamma_4 x\sqrt{\tilde{\tau}_4}\right)}{\vo^2} \\
^{\prime \prime } & ^{\prime \prime } & \frac{\alpha \gamma_3}{\vo\sqrt{\tau_3}}
& \frac{3\alpha^2 \gamma_3\gamma_4\sqrt{\tau_3\tilde{\tau}_4}}{\vo^2} &
\frac{3\alpha^2\gamma_3\sqrt{\tau_3}\left(\gamma_5\sqrt{\tau_5}-\gamma_4 x\sqrt{\tilde{\tau}_4}\right)}{\vo^2} \\
^{\prime \prime } & ^{\prime \prime } & ^{\prime \prime } &
\frac{\alpha \gamma_4}{\vo\sqrt{\tilde{\tau}_4}} &
-\frac{\alpha \gamma_4 x}{\vo\sqrt{\tilde{\tau}_4}} \\
^{\prime \prime } & ^{\prime \prime } & ^{\prime \prime } &
^{\prime \prime } & \frac{\alpha}{\vo}\left(\frac{\gamma_4 x^2}{\sqrt{\tilde{\tau}_4}}
+\frac{\gamma_5}{\sqrt{\tau_5}}\right)
\end{array}
\right) ,  \label{LaDiretta}
\end{equation}
where, as in \cite{fiberinfl}, we have systematically dropped terms that
are suppressed relative to the ones shown by factors
$\sqrt{\tau_i/\tau_2}$ $\forall\,i=3,4,5$.

It is convenient to trade $\tau_2$ for $\vo$ and $\tau_4$ for
the combination $\tilde{\tau}_4\equiv \tau_4-x\tau_5$.
In the limit in which $\tau_1$, $\tau_2$ are much larger
than $\tau_3$, $\tau_4$ and $\tau_5$, the kinetic Lagrangian can be canonically normalised
order by order in $1/\vo$ writing:
\be
 {\cal L}_{kin} = {\cal L}_{kin}^{\mathcal{O}(1)}+{\cal
 L}_{kin}^{\mathcal{O}(\vo^{-1})} \,, 
\ee
where the leading term is\footnote{We use units with $8 \pi M_p = 1$ unless
otherwise stated.}
\be
 - \frac{{\cal L}_{kin}^{\mathcal{O}(1)}}{\sqrt{-g}}
 =\frac{3}{8 \tau_1^2}\left( \partial \tau_1\right)^2
 +\frac{1}{2 {\cal V}^2}\left( \partial
 {\cal V}\right)^2 - \frac{1}{2 \tau_1 {\cal V}}
 \,\partial \tau_1 \partial {\cal V} \,,
 \label{LagKinO(1)}
\ee
while the subleading term at $\O(1/\V)$ reads
\bea
 - \frac{{\cal L}_{kin}^{\mathcal{O}(\vo^{-1})}}{\sqrt{-g}}
 &=&\frac{3 \alpha }{8\vo} \left[\frac{\gamma_3}{\sqrt{\tau_3}} \left(\partial\tau_3 \right)^2
+\frac{\gamma_4}{\sqrt{\tilde\tau_4}} \left(\partial\tilde\tau_4 \right)^2
 +\frac{\gamma_5}{\sqrt{\tau_5}}\left(\partial\tau_5\right)^2 \right] \nonumber \\
&-&\frac{3 \alpha }{4\vo}\frac{\partial\tau_1}{\tau_1}\left[\gamma_3\sqrt{\tau_3}\partial\tau_3
 +\gamma_4\sqrt{\tilde\tau_4} \partial\tilde\tau_4 +\gamma_5\sqrt{\tau_5}\partial\tau_5\right].
 \label{LagKinO(V-1)}
\eea
At $\mathcal{O}(1)$ the transformation
\bea
 \tau_1 &=& \exp{\left(a \phi + b \chi_2\right)},
 \label{tau1}\\
 \vo &=& \exp{\left(c \chi_2\right)} \,, \label{vol}
\eea
puts the expression (\ref{LagKinO(1)}) into canonical form
\be
 - \frac{{\cal L}_{kin}^{\mathcal{O}(1)}}{\sqrt{-g}}
 =\frac{1}{2}\left[\left( \partial \phi\right)^2
 +\left( \partial \chi_2\right)^2 \right] \,,
\ee
where the coefficients $a$, $b$ and $c$ are obtained from the
condition that the matrix $M = \left( \begin{array}{cc} a & b \\ 0
& c \end{array} \right)$ satisfies
\be
    M^{\scriptscriptstyle T} \cdot
 \left( \begin{array}{rrr}
  \frac34 && -\frac12 \\
  -\frac12 && 1 \\
 \end{array} \right) \cdot M
 = I \,.
\label{condition}
\ee
This has four solutions: $(a,b,c)$, $(a,-b,-c)$, $(-a,b,c)$, $(-a,-b,-c)$, where:
\begin{eqnarray}
 a &=&\frac{2}{\sqrt{3}} ,\text{ \ \ \ \ \ \ \ }
 b = \sqrt{\frac23} ,\text{ \ \ \ \ \ \ \ }
 c =\sqrt{\frac32}\, ,
\end{eqnarray}
and for concreteness we shall choose the first one will all plus signs.
As is shown in \cite{Reheating}, the fields $\chi_1$ and $\chi_2$ turn
out to also diagonalise the mass-squared matrix, $M^2_{ij} =\sum_k
K^{-1}_{ik}V_{kj}$ in the limit where string-loop corrections to
the potential are neglected. Once string-loop corrections are
included a subdominant dependence of $\cal V$ on $\chi_1$ also
arises and (\ref{vol}) gets modified to:
\be
\vo=\exp{\left(\sqrt{\frac{3}{2}}\chi_2+\kappa\left(\phi,\chi_2\right) \phi\right)},
\label{volloop}
\ee
where $\kappa\left(\phi,\chi_2\right)\sim\mc{O}\left(\vo^{-1/3}\right)$
is a subleading correction which in the limit of exponentially large volume and
large $\tau_1$ takes the form:
\be
\kappa\left(\chi_1,\chi_2\right)=\sqrt{2}\left(\frac{2}{3}\right)^{15/4}\frac{b_3^{3/2}}{\sqrt{\chi_2}}\,e^{-\frac{1}{\sqrt{6}}\chi_2}
\left(B \,e^{-\frac{1}{\sqrt{3}}\phi}-4 A\,e^{-\frac{4}{\sqrt{3}}\phi}\right).
\ee
Next we diagonalise the next-order kinetic term,
$\mathcal{L}_{kin}^{\mathcal{O}(\vo^{-1})}$. The first line in
(\ref{LagKinO(V-1)}) becomes diagonal once we rescale the three
small moduli as follows
\be
 \tau_3=\left(\frac{3 \vo}{4 \alpha \gamma_3} \right)^{2/3} \chi_3^{4/3}, \quad
 \tilde\tau_4=\left(\frac{3 \vo}{4 \alpha \gamma_4} \right)^{2/3} \chi_4^{4/3}, \quad
 \tau_5=\left(\frac{3 \vo}{4 \alpha \gamma_5}  \right)^{2/3} \sigma^{4/3}\,.
 \label{smalltaus}
\ee
Notice that the original modulus $\tau_4$ is a combination of $\chi_4$ and $\chi_5$
since (\ref{smalltaus}) implies that
\be
\tau_4=\tilde\tau_4+x\tau_5=\left(\frac{3 \vo}{4 \alpha}
 \right)^{2/3}\left(\frac{\chi_4^{4/3}}{\gamma_4^{2/3}}+x\frac{\sigma^{4/3}}{\gamma_5^{2/3}}\right).
\label{CanNormTau4}
\ee
The second line in (\ref{LagKinO(V-1)}) is similarly diagonalised by mixing
$\tau_1$ with $\tau_j$ $\forall\,j=3,4,5$. Explicitly, introducing the
following subleading corrections to (\ref{tau1}):
\be
 \tau_1 = \exp{\left(\frac{2}{\sqrt{3}}
 \, \phi + \sqrt{\frac23} \, \chi_2
 + \frac{1}{2}\sum_{j=3}^4 \chi_j^2 + \frac{1}{2} \sigma^2 \right)}\,, \label{Tau1}
\ee
gives to this order:
\be
 {\cal L}_{kin}^{\mathcal{O}(1)}
 +{\cal L}_{kin}^{\mathcal{O}(\vo^{-1})}
 =\frac{1}{2}(\partial\phi)^2 + \frac12\sum_{i=2}^4\left(\partial \chi_i\right)^2
 +\frac{1}{2}\left(
 \partial \sigma\right)^2 \,.
\ee
Notice that the last two terms in eq.~(\ref{Tau1}) are
subleading because $\chi_j \sim \sigma \sim \O(\vo^{-1/2}) \ll 1$ for $j=3,4$,
while from (\ref{tau1}) and (\ref{vol}), we have $\phi\sim\chi_2 \sim
\O(\ln\vo)$. We can now substitute (\ref{volloop}) in
(\ref{smalltaus}) to eliminate $\V$ and directly express $\tau_i$, $i=3,4,5$,
in terms of the canonically normalised fields:
\bea
&& \tau _{3}\, \simeq\,\left( \frac{3}{4\alpha
 \gamma_{3}}\right) ^{\frac{2}{3}}
 \exp {\left(\sqrt{\frac23} \,\chi_2+ \frac{2}{3}\kappa\left(\phi,\chi_2\right)\phi   \right) }
 \,\chi_{3}^{\frac{ 4}{3}},  \nonumber \\
 && \tau _{5}\, \simeq\,\left( \frac{3}{4\alpha
 \gamma_{5}}\right) ^{\frac{2}{3}}
 \exp {\left(\sqrt{\frac23} \,\chi_2+ \frac{2}{3}\kappa\left(\phi,\chi_2\right)\phi \right) }
 \,\sigma^{\frac{ 4}{3}},
\eea
and
\be
 \tau_4=\left( \frac{3}{4\alpha}\right)^{\frac{2}{3}}
 \exp {\left(\sqrt{\frac23} \,\chi_2+ \frac{2}{3}\kappa\left(\phi,\chi_2\right)\phi\right) }
 \left(\frac{\chi_4^{4/3}}{\gamma_4^{2/3}}+x\frac{\sigma^{4/3}}{\gamma_5^{2/3}}\right).
 \label{t4CN}
\ee
Notice that we have neglected subleading mixing terms between the blow-up modes which
will be introduced by subdominant loop corrections.
The field redefinitions we have determined
render canonical the form of the kinetic terms.

\section{Moduli couplings to gauge bosons}
\label{app-parameters}

In this section, we write the expressions for the parameters $u_i$
introduced in section \ref{couplings}, when calculating the moduli couplings to visible sector gauge
bosons. They read (here the VEV of the dilaton is simply $\langle s\rangle = g_s^{-1}$):
\bea
u_0 &\equiv& 1 +\frac{ h\left(\mc{F}_{\rm vis}\right) \langle s\rangle}{\langle\tau_{\rm vis}\rangle+\hat{\tau}_{\rm vis}} =
1 +\frac{ h\left(\mc{F}_{\rm vis}\right) \,g_s^{-1}}{v_4 \langle\tilde\tau_4\rangle
+\left(x v_4+v_5\right)\langle\tau_5\rangle\left(1+x_{\sigma}\right)^{4/3}}\,, \notag \\
u_1 &=& \frac{1}{9} \left(\frac{2}{3}\right)^{\frac{13}{4}} \frac{b_3^{3/2}e^{-\frac{\langle\chi_2\rangle}{\sqrt{6}}}}{\sqrt{\langle\chi_2\rangle}}
\left[16 A \left(\langle\phi\rangle-\frac{\sqrt{3}}{4}\right) e^{-\frac{4 \langle\phi\rangle}{\sqrt{3}}}-B
\left(\langle\phi\rangle-\sqrt{3}\right) e^{-\frac{\langle\phi\rangle}{\sqrt{3}}}\right], \notag \\
u_2 &=& 1+\frac{1}{3\sqrt{3}}\left(\frac{2}{3}\right)^{\frac{15}{4}}\frac{b_3^{3/2}\langle\phi\rangle}{\langle\chi_2\rangle^{3/2}}
\left(\sqrt{2} \,\langle\chi_2\rangle+\sqrt{3}\right) e^{-\frac{\langle\chi_2\rangle}{\sqrt{6}}}
\left(4 A \,e^{-\frac{4 \langle\phi\rangle}{\sqrt{3}}}
-B e^{-\frac{\langle\phi\rangle}{\sqrt{3}}}\right), \notag \\
u_4&=& \left[\langle\chi_4\rangle
 +\frac{\left(x\,v_4+v_5\right)}{v_4} \left(\frac{\gamma_4}{\gamma_5}\right)^{2/3}\left(\langle\sigma\rangle+\hat\sigma\right)
 \left(\frac{\langle\sigma\rangle+\hat\sigma}{\langle\chi_4\rangle}\right)^{1/3}\right]^{-1}, \notag \\
u_5&=& \left[\langle\sigma\rangle+\hat\sigma+\frac{v_4}{\left(x\,v_4+v_5\right)}\left(\frac{\gamma_5}{\gamma_4}\right)^{2/3} \langle\chi_4\rangle
  \left(\frac{\langle\chi_4\rangle}{\langle\sigma\rangle+\hat\sigma}\right)^{1/3}\right]^{-1}.
\notag
\eea

\end{appendix}

\end{document}